


\def\di{\partial}

\def\lix{${\it \$}$_\xi}
\def\Lix{\hbox{{\it \$}}_\xi}


\def\sig{\sigma}

\def\bxi{{\bar\xi}}

\def\lam{\lambda}

\def\Del{\Delta}
\def\del{\delta}
\def\Bg{\Bar g}
\def\hg{\hat g}
\def\k{\kappa}

\def\Ga{\Gamma}


\def\~{\tilde}
\def\g.mn{g_{\mu\nu}}

\def\half{{\textstyle{1 \over 2}}}

\def\xm{x^\mu}
\def\f.kl{f_{kl}}

\def\lag{\hat{\cal L}}

\def\m{\eqno}
\def\l{\left}
\def\r{\right}

\def\txt{\textstyle}

\def\Bar{\overline}


\def\bbb{background~}
\def\bbbb{backgrounds~}

\def\cccc{conservation laws~}
\def\ckkk{conformal Killing vector~}
\def\ckkkk{conformal Killing vectors~}
\def\ee'e{Einstein's equation~}
\def\ee'ee{Einstein's equations~}

\def\em{energy-momemtum~}
\def\frw{Friedmann-Robertson-Walker~}

\def\frwwww{Friedmann-Robertson-Walker spacetimes~}
\def\gr{general relativity~}
\def\kkk{Killing vector~}
\def\kkkk{Killing vectors~}
\def\nnn{\noindent}

\def\sss{superpotential~}
\def\ssss{superpotentials~}
\def\spttt{spacetime~}
\def\sptttt{spacetimes~}


\magnification=\magstep 1

\def \singlespace {\baselineskip = 14pt plus 3pt \message {single space}}
\singlespace

\centerline{\bf  RELATIVISTIC CONSERVATION LAWS ON CURVED BACKGROUNDS}
\centerline{{\bf AND THE THEORY OF COSMOLOGICAL PERTURBATIONS}\footnote
*{A summary of this work
has been presented at
the Conference on
``Fundamental Interactions: From Symmetries to Black
Holes'', Brussels, March 25-27, 1999. It has been published in the
Proceedings of the Conference
under the  title {\it Conservation Laws for Large
Perturbations on Curved Backgrounds} with ref.:
{\it Fundamental Interactions: From Symmetries to Black
Holes} (eds.: J.M. J.M. Fr\'ere, M. Henneaux, A.Servin \& Ph. Spindel),
p.p. 147-157, Brussels: Universit\'e Libre de Bruxelles (1999),
see also gr-qc/9905088.}}
\vskip .5 in

\centerline{Alexander N. Petrov\footnote\dag{E-mail: petrov@sai.msu.su}}

\centerline{Sternberg Astronomical Institute, Universitetskii prospect 13,
Moscow 119899, Russia}

\centerline{and}

\centerline{Joseph Katz\footnote\ddag{E-mail: jkatz@vms.huji.ac.il}}

\centerline{The Racah Institute of Physics, 91904 Jerusalem, Israel}

\vskip .5 in

\centerline{ABSTRACT}

\vskip .3 in

We first consider the Lagrangian formulation of general relativity for
perturbations with respect to a background spacetime. We show that by
combining N\oe ther's method  with Belinfante's ``symmetrization''
procedure we obtain conserved vectors that are independent of any
divergence added to the perturbed Hilbert Lagrangian. We also show that
the  corresponding perturbed energy-momentum tensor is symmetrical and
divergenceless but only on backgrounds that are ``Einstein spaces'' in
the sense of A.Z. Petrov.  de Sitter or anti-de Sitter and Einstein
``spacetimes" are Einstein spaces but in general
Friedmann-Robertson-Walker spacetimes are not.  Each conserved vector
is a divergence of an anti-symmetric tensor, a ``superpotential". We find
superpotentials which are a generalization of Papapetrou's
superpotential and are rigorously linear, even for large perturbations,
in terms of the inverse metric density components and their first order
derivatives. The \ssss  give correct globally conserved quantities at
spatial infinity. They resemble Abbott and Deser's superpotential, but
give correctly the Bondi-Sachs total four-momentum at null infinity.

Next we calculate conserved vectors and \ssss for perturbations of a
Fried-mann-Robertson-Walker \bbb associated with its 15  \ckkkk given in
a convenient form.  The integral of each conserved vector in a
finite volume
$V$ at a given conformal time is equal to a surface integral on the
boundary of $V$ of the superpotential.  For given boundary conditions
each such integral
is part of a flux whose total through a closed
hypersurface is equal to zero. For given boundary conditions on $V$, the
integral can be considered
as an ``integral constraint" on data in the volume and
this data  always includes the energy-momentum perturbations.
We give explicitly
these 15 integral constraints and add some simple applications  of
interest in cosmology. Of particular interest are Traschen integral
constraints in which the volume integral contains only the matter
energy-momentum tensor perturbations and not the field perturbations.
We show that these particular integral constraints are associate with
time dependent linear combinations of conformal Killing vectors. Such
linear combinations are neither \kkkk nor conformal Killing vectors. We
also find that if we add the ``uniform Hubble constant hypersurface"
gauge condition of Bardeen, there exists 14 such integral constraints.
The exception is associated with conformal time translations ($k=\pm
1$) or conformal
time accelerations ($k=0$).  As an example we find the constants
of motion of a \spttt that is asymptotically Schwarzschild-de Sitter
($k=0$).

\beginsection  1. Introduction

\noindent (i) {\it Conservation laws and cosmology.}

Conservation laws associated with ``symmetric'' infinitesimal
displacements in Fried- man-Robertson-Walker spacetimes have been used
in relativistic cosmology on several occasions. Infinitesimal
displacements  are characterized by vector fields and, as we shall see,
the vectors used in some applications  are not always \kkkk nor even
conformal Killing vectors.

An  example in which no Killing nor \ckkkk are used has been given by
Traschen [1] who introduced ``integral constraints'' in terms of
``integral constraints vectors''.  Traschen and Eardley [2]  analyzed
measurable effects of the cosmic background radiation due to spatially
localized perturbations.  By using ``integral constraints'' they
pointed to an important reduction of the Sachs-Wolfe [3] effect on the
mean square angular fluctuations at large angles of the cosmic
background temperature due to local inhomogeneities. Traschen's
integral constraints vectors have a somewhat intriguing origin [4].
The equations for integral constraint vectors have been studied   by
Tod [5]. He showed that these equations are conditions for a spacelike
hypersurface to be embeddable in a \spttt with constant curvature of
which the  solutions are Killing vectors. In Katz, Bi\u{c}\'ak and
Lynden-Bell [6], a paper  referred to as KBL97, integral constraints
appear as conservation laws with \kkkk in a de Sitter background; more
on this below.

Local differential conservation laws, rather than global ones, have
been used by Veeraraghavan and Stebbin [7]. They found and used a
conserved ``energy-momentum" pseudo-tensor in an effort to integrate
Einstein's equations with scalar perturbations and topological defects
in the limit of long wavelengths on a Friedmann-Robertson-Walker \spttt
with flat spatial sections $(t=const, k=0$). Uzan, Deruelle and Turok
[8] realized that these conservation laws might be associated with the
conformal  Killing vector of time translations  and they
extended Veeraraghavan and Stebbin's  method
to \frw perturbed spactimes with non-flat spatial
sections $(k=\pm 1)$. More on this in section 4.

In Lynden-Bell, Katz and Bi\v c\'ak's [9] study of Mach's principle
from the relativistic constraint equations,   conservation laws yield a
general proof that the total angular momentum (and the total of any
conserved perturbation of the current which deriving from a
``superpotential'') must be zero in any closed universe.

As a final example  we mention  KBL97's  analysis of the globally
conserved quantities that result from mapping a \frw perturbed \spttt
on a de Sitter background with its ten Killing vectors.

With these different examples in mind, it made  good sense to study the
properties and physical interpretation  of conservation laws and their
superpotentials, in the context of relativistic cosmology, associated
with arbitrary displacements in a \bbb as was done in KBL97. In fact,
the theory has wider applicability than relativistic cosmology since
the \bbb may be any \spttt and there are plenty of examples
in general relativity in which
\bbbb are used.

\nnn (ii) {\it N\oe ther's method and its problems.}

KBL97 used  the fairly standard method of  N\oe ther (see for instance
Landau and Lifshitz [10]) to derive \cccc from the Lagrangian $ \lag_G$
of the perturbations of the gravitational field\footnote{**} {Notations
are properly defined in section 2. Here we assume the reader to be
familiar with current notations.}
$$
\lag_G  =\lag  -\Bar {\lag},~~~~ \lag= -{1\over 2\k}(\hat R + \di_\mu
\hat k^\mu),~~~~\Bar{\lag}= -{1\over 2\k}\Bar {(\hat R + \di_\mu
\hat k^\mu)},~~~~\k = {{8\pi G}\over
{c^4}}.
\m (1.1)
$$
Here a $~\hat{}~$ means multiplication by $\sqrt {-g}$, a bar referees
to the background, R is the scalar curvature and $\hat k^\mu$ is some
vector density.  N\oe ther's method associates a conserved vector
density $\hat I^\mu$ with {\it any} vector $\xi^ \mu$ that generates
small displacements. It applies to perturbation theory on {\it any}
background and provides a ``canonical'' energy-momentum tensor
perturbation. Moreover, the conserved vector is always [11] the
divergence of an antisymmetric tensor, a superpotential, ($\hat I^\mu=
\di_\nu \hat I^{\mu\nu}, ~\hat I^{\mu\nu}=-\hat I^{\nu\mu} $).  This
has the great practical advantage that  integrals of  complicated $\hat
I^\mu$'s  in a volume $V$ are equal to often much simpler integrals of
$\hat I^{\mu\nu}$'s on the boundary of $V$.  N\oe ther's method is the
most direct and easy way to construct superpotentials, field energy
tensors and conserved vector densities with arbitrary backgrounds and
for arbitrary $\xi^ \mu$ though the same results can of course be
worked out directly from the perturbed Einstein equations.  But, at
least in our case, this is far more complicated than with the method
developed here as we shall see.

N\oe ther's method has, however, two unsatisfactory features. First the
Lagrangian density is not unique. A divergence $\di_\mu\hat k^\mu$  can
and must be added to the Hilbert Lagrangian because the latter leads [12] to
Komar's [13] conservation law which gives the wrong mass to
angular momentum ratio with an ``anomalous'' factor of two in the weak
field limit [14] and  does not give the Bondi mass [15]  at null
infinity [16]. Divergences are also added to comply with different
boundary conditions. Various divergences have thus been added to
$\hat R$ for different reasons. M\o ller [17], using a tetrad
representation $e^\alpha_\mu$ with $g_{\mu\nu} =\eta_{\alpha\beta}
e^\alpha_\mu e^\beta_\nu$, would have taken a $\hat k^\mu = \hat
g^{\mu\nu}g^{\rho\sig}(e^\alpha_\rho D_\sig e^\beta_\nu - e^\alpha_\nu
D_\rho e^\beta_\sig) \eta_{\alpha\beta}$.  York [18], using a
foliation, would have chosen
$\hat k^\mu = 2\l( \varepsilon n^\mu D_\nu \hat n^\nu -
n^\nu D_\nu \hat n^\mu\r)$  with
$n^\mu$  ($n^\mu n_\mu = \varepsilon =\pm 1$) the
normal vectors of his closed hypersurfaces.  KBL97 wanted a field
energy tensor quadratic in first order derivatives and took therefore
like Rosen [19] a long time before
$$
\hat k^\mu = {1\over \sqrt{-g}}
\Bar D_\nu (-gg^{\mu\nu}).
\m(1.2)
$$
$\Bar D_\mu$ is a covariant derivation with respect to the \bbb metric
$\bar g_{\mu\nu}$.  Second, the canonical field energy momentum tensor
is not symmetrical nor is it divergenceless. On a flat background, the
energy-momentum tensor {\it is} divergenceless but is still not
symmetrical and the angular momentum is not conserved; it does not
include the helicity of the field. It thus appears that conservation
laws obtained by N\oe ther's method have an unsatisfactory weak field
limit on a flat background at least as far as angular momentum is
concerned.

To remedy that situation we suggest in this paper to modify  N\oe ther
conserved vectors using Belinfante's [20] trick in classical field
theory. It is an easy matter to adapt his method to perturbation theory
on curved backgrounds.  Belinfante's modification leads to
energy-momentum tensors which  ensures, at least in classical field
theory, that  angular momentum includes the helicity and is then
conserved.  The Belinfante trick has  been applied  by Papapetrou [21]
to \gr in an effort to calculate the total angular momentum at spatial
infinity.

These new conserved vectors have none of  the drawbacks just described
and in addition  have very appealing new properties.

\noindent (iii) {\it A summary of  theoretical results.}

It may be useful, at this stage, to give a summary of our main
theoretical results, we mean those valid in general, not only in
relativistic cosmology.

\noindent (a) We find that there exist a conserved vector density $\hat
{\cal I}^\mu$ associated with any vector generating infinitesimal
displacements and which is the divergence of a superpotential
$\hat {\cal I}^{\mu\nu}$;  it is of the following form
$$
\hat {\cal I}^\mu =\hat {\cal T}^\mu_\nu \xi^\nu + \hat {\cal Z}^\mu
= \di_\nu{\hat
{\cal I}^{\mu\nu}}.
\m(1.3)
$$
$\hat {\cal T}^\mu_\nu$ represents a matter plus field energy-momentum
tensor   density perturbation relative to the background.  $\hat {\cal
Z}^\mu$ is  a vector density that is only equal to zero if $\xi^\nu$ is
a Killing vector of the background, which we denote then by $\bar
\xi^\nu$. As a consequence of Eq. (1.3) a volume integral $F$ of $\hat
{\cal I}^\mu$ equal the surface integral of $\hat {\cal I}^{\mu\nu}$ on
its boundary $S$
$$
F = \int_V  \hat {\cal ~I^\mu} dV_\mu = \oint_S \hat {\cal I}^{\mu\nu}
dS_{\mu\nu}
\m(1.4)
$$
and the total flux $F$ through a closed hypersurface V is equal to
zero.  That is what is meant $\hat {\cal I^\mu}$ being a conserved
current.  Equalities such as (1.4) may be regarded as ``integral
constraints" in the following sense. Suppose that  boundary values on
$S$ and thus $ \hat {\cal I}^{\mu\nu}$ are known. Then Eq. (1.4)
represents constraints on the perturbations of the energy-momentum
tensor which is always part of $\hat {\cal T}^\mu_\nu$. There are thus
as many integral constraints as there are displacement vectors
$\xi^\mu$ (but not all are equally interesting).  Our notion of
``integral constraints" is slightly different from that introduced by
Traschen [1]. She calls integral constraint an expression like Eq.
(1.10) below in which the boundary term is equal to zero and the volume
integral contains only the perturbation of the matter energy-momentum
tensor.
One may wish  to see in Eq. (1.4) a definition of quasi-local
conservation laws if the boundary is not at the border of \spttt
itself.

\nnn (b) The conserved vector, and thus the corresponding $\hat {\cal
T}^\mu_\nu$ are {\it independent} of any divergence added to
the Hilbert Lagrangian density of the perturbations. It has been
noticed before by Bak, Cangemi and Jackiw [22]  that Belinfante's
modification of the N\oe ther currents obtained from Hilbert's or
Einstein's Lagrangians lead to the same symmetric and divergenceless
energy-momentum tensor relative to a flat background in Minkowski
coordinates. Ours is a generalization of this finding for any
divergence added to the Hibert Lagrangian, for arbitrary
perturbations with
respect to any background in arbitrary coordinates.

We want to stress that since our conserved vectors are independent of
an added divergence, they are also  {\it independent} of boundary
conditions. This result is in line with classical field ideas. The
opposite view that pseudotensors and superpotentials must depend
on boundary conditions has been held for instance in [23].

\noindent (c) From Eq. (1.3) follows that for each Killing vector of the
background
$\bxi^\nu$ there exists a conserved vector $\hat {\cal J}^\mu$ and a
corresponding
superpotential
$\hat
{\cal J}^{\mu\nu}$ such that
$$
 \hat {\cal J}^\mu =\hat {\cal T}^\mu_\nu \bxi^ \nu=\di_\nu{\hat
{\cal J}^{\mu\nu}}.
\m(1.5)
$$
This expression looks also very much like a conservation law in
classical field theory.

\noindent (d)  $\hat {\cal T}^{\mu\nu}=\hat {\cal T}^\mu_\rho \Bar
g^{\rho\nu}$ is {\it symmetrical} and {\it divergenceless} if and only if
the
background is an Einstein space in the sense of A.Z. Petrov [24], that
is if
$
\Bar R_{\mu\nu} = - \Bar{ \Lambda} \Bar g_{\mu\nu}
$
where  $\Bar \Lambda$ is necessarily a constant.  de Sitter spacetimes
belong to that category.  Other Friedmann-Robertson-Walker spacetimes
that are currently used in cosmology do not.

\noindent (e) The new superpotential  $~\hat {\cal I}^{\mu\nu}$ is
reminiscent of many well known ones (see section 2 for details) and
has a simple form:
$$
\hat {\cal I}^{\mu\nu} =  -\hat
{\cal I}^{\nu\mu}={1 \over \k} \hat l^{\sig[\mu}\Bar D_\sig
\xi^{\nu]}
+   {1 \over \k}\Bar D_\sig \l(\hat l^{\rho[\mu}\Bar
g^{\nu]\sig} - \hat l^{\sig[\mu} \Bar g^{\nu]\rho}\r) \xi_{\rho}.
\m(1.6)
$$
In this expression $\hat l^{\mu\nu}$ is the perturbed inverse metric density:
$$
\hat l^{\mu\nu} =\hat g^{\mu\nu} -\Bar {\hat g^{\mu\nu}}.
\m (1.7)
$$
The \sss has the remarkable property of being {\it linear} in $ \hat
l^{\mu\nu}$.  Linearity is a valuable property; the linear
approximation is not different from the non linear one. Global exact
conservation quantities of know asymptotic fields with unknown sources can be
calculated and given physical meaning.  The superpotential (1.6) satisfies
standard criteria of global conservation laws in asymptotically flat
spacetimes at spatial and at null infinity (see appendix).  Also
second order corrections of the energy-momentum tensor due to field
energy contributions are readily calculable from our formulas.

Those are the principal theoretical results of the paper.

\noindent (iv) {\it The 15 Conformal Killing vectors of cosmological
backgrounds and their associated conservation laws and integral
constraints. }

To illustrate our new conservation laws in theoretical cosmology we
consider the conserved vectors and \ssss associated with the 15 \ckkkk
of \frw spacetimes. By \ckkkk we mean the 15 linearly independent
solutions of the \ckkk equations
$$
\Bar D_{(\mu} \xi_{\nu)}  = {\textstyle{1\over 4}}\bar g_{\mu\nu}
\Bar D_\rho \xi^\rho,~~~~~\xi_\nu=\bar
g_{\nu\rho}\xi^\rho.
\m (1.8)
$$
The 15 \ckkkk include by definition the 6 pure \kkkk for which $\Bar
D_\rho \xi^\rho=0$. \frwwww are conformal to Minkowski's spacetime. There
are thus similarities  between the 4 \kkkk of  translations, the 3
rotations, 3 Lorentz boosts, 3 center of mass position, 1 dilatation
and the 4 ``accelerations'' of Minkowski's spacetime. Such similarities are
helpful in geometrical interpretations. A presentation of those \ckkkk
of Minkowski \spttt in a form that appeals to physicists is given in
Fulton, Rohrlich and Witten [25] who have a particular liking for the
accelerations which they explain well and to which we referee the reader
interested in those slightly unfamiliar \ckkkk.

Now follows a brief summary of the results.

\nnn (a) We give the 15 linearly independent solutions of Eq. (1.8) in
a simple mathematical form. We are not interested in the algebra of the
conformal group. We are mostly interested here in quasi-local or in
global conservation laws or integral constraints for volumes in a
sphere (parametrized by $r$) at a given instant of conformal time $\eta$.
Thus, in
$(\eta,x^k)$ coordinates we are looking for integrals of the form
$$
\int_\eta  {\cal I}^0 dV = \oint_r {\cal I}^{0l} dS_l.
\m(1.9)
$$
Any linear combination of such integrals with time dependent
coefficients are still of the same form. We give what seem to us the 15
simplest linear combinations. Notice that linear combinations of \ckkkk with
time dependent coefficients are not \ckkkk anymore.
Notice also that the
left hand integrand of such linear combinations is not a zero component of
a conserved vector anymore.

\nnn (b) We next turn our attention to those linear combinations in
which the volume integrand depends only on the matter \em perturbations
$\delta T^0_\mu$ thus of the form
$$
\int_V  \delta T^0_\nu V^\nu dV = \oint_S  B^l dS_l.
\m(1.10)
$$
There are 10 integral constraints of this form, 6 are associated with
the pure \kkkk of the \frwwww and 4 are the vectors found by Traschen
[1]. Thus Traschen's ``integral constraint vectors" appear here as
linear combinations of \ckkkk with time dependent coefficients.

\nnn (c) We then show that if we apply the uniform Hubble expansion
gauge studied by Bardeen [26] all but one of the 15 \ckkkk are
associated with integral conservation laws of the form (1.10). The
exception is associated with conformal time translations if $k=\pm 1$
or  conformal time accelerations if $k=0$.
A look at Eq. (1.10) shows that these
integrals might be constructed  directly from Einstein's constraint
equations. However the $V^\mu$'s though simple are not all that easy to
guess as we shall see. The integral constraints have often simple
geometrical interpretations by analogy with classical mechanics.
With \ckkkk they
are momenta of order 0, 1 or 2.

\nnn (d) We also give a non-trivial example of globally conserved
quantities on a \bbb that is not asymptotically flat. We calculate the
constants of motion of \sptttt that are asymptotically Schwarzschild-de
Sitter (with k=0). Because of the high degree of symmetry of the
asymptotic conditions we find 13 globally conserved quantities equal to
zero, that is 13 Traschen's like integral constraints. There are 2
constants of motion which are different from zero.

Such  are the examples of  cosmological interest studied in this paper.

\noindent (v) {\it Presentation of the paper.}

In the following section we describe in detail the way to obtain N\oe
ther's conservation laws on curved backgrounds. A few parts of that
chapter are taken from KBL97 but they are  well worth repeating here.
Section 3 introduces the Belinfante correction to N\oe ther's conserved
vectors. There we show that the modified conserved vectors
and their associated superpotentials are
unchanged if we add a divergence to the Hilbert Lagrangian for the
perturbations. At the end of section 3 we derive Rosenfeld's [27]
identities which give beautiful relations among complicated quantities
of interest.
Rosenfeld's identities are refereed to as ``cascade
equations" by Julia and Silva [28].
In that section we also obtain the theoretical results
summarized above in (iii).  In section 4 we briefly describe how we
found  the 15 \ckkkk of \frwwww in appropriate coordinates and how to
calculate  the corresponding conserved vectors and superpotentials in a
$1+3$ standard decomposition. The same section contains the examples
just described in (iv).  In section 5 we emphazise the role of
superpotentials in the development of conservation law theory for
general relativity and make a short review of historically important
\ssss on a flat background.
A flat \bbb is quite useful to study conservation
laws in general relativity as pointed out by Rosen [19] a long time ago.

Finally in appendix we briefly show that our new superpotential has
the normally expected global properties at null and spatial infinity in
asymptotically flat spacetimes.

Each section is preceded by a summary which gives the motivations and
point out where the reader will find the principal formulas. The body
of the sections themselves are  written for  readers who are interested
in the mathematical details.  However, most elaborate but
straightforward calculations are not given in detail.
Unfortunately \gr is replete with them.

\vskip .1 in

\beginsection {2. N\oe ther}

(i) {\it Motivations and summary of results.}

In this section we apply N\oe ther's method to the Lagrangian $\lag_G$
defined in Eq. (1.1) with the vector density $\hat k^\mu$ defined in
Eq.  (1.2).  This is the KBL97 Lagrangian. The conserved vector $\hat
I^\mu$ obtained in this way is given in Eq.  (2.17). A look at Eq.
(2.17) shows that it contains a canonical energy-momentum tensor
density $ \hat \theta^\mu_\nu$ explicitly written in Eq. (2.18) with
$\hat t^\mu_\nu$ given in Eq. (2.13), an ``helicity'' term $\hat
\sig^{\mu\rho\sig}$ given by Eq. (2.14) and a vector density
$\hat\eta^\mu$, see Eq.  (2.19), defined in terms of the {\it
derivatives} of the Lie derivatives of the background
metrics $\bar g_{\rho\sig}$ or $\bar
z_{\rho\sig}$ defined in Eq. (2.11); $\hat\eta^\mu$ is thus zero if
$\xi^\mu$ is a Killing vector $\bar \xi^\mu$ of the background.

The canonical energy-momentum tensor density is neither symmetrical nor
divergenceless except on a flat background [see Eq. (5.1{\it a})] in which
case the canonical field energy-momentum $\hat t^\mu_\nu$ reduces to
Einstein's pseudo-tensor  in Minkowski coordinates $X^\mu$. In the
same coordinates the helicity tensor density $\hat
\sig^{\mu[\rho\sig]}$ is that which was given by Papapetrou [21].
The conserved vector
density $\hat I^\mu$ is thus a generalization of standard old results
to finite perturbations of a curved background with arbitrary vectors
not only with Killing vectors. $\hat I^\mu$ is also written as the
divergence of a superpotential $\hat I^{\mu\nu}$ which is given in Eq.
(2.21) and also by Eq. (2.22).

The present section also contains the energy-momentum tensor,
the helicity, and
the superpotential that are obtained from the Hilbert Lagrangian density
$$
\lag'_G= -{1\over 2\k}(\hat R -\Bar {\hat R}).
\m (2.1)
$$
The quantities, $ \hat\theta'^\mu_\nu$ and  $\hat \sig'^{\mu\rho\sig}$
are defined   by Eqs. (2.24);  the superpotential $\hat K^{\mu\nu}$
is given by
``Eq. (2.21) minus its $\xi k$-term". While
$\hat\theta^\mu_\nu$ and, to a certain extend, also  $\hat
\sig^{\mu\rho\sig}$ reduce to familiar quantities on a flat background,
$ \hat\theta'^\mu_\nu$ and  $\hat \sig'^{\mu[\rho\sig]}$ are much more
complicated; $\hat\theta'^\mu_\nu$ contains  second order
derivatives. In fact we gain in clarity  and simplicity by starting the
calculations with $\lag_G$ rather than $\lag'_G$; the end product
in section 3 is independent of $\di_\mu \hat k^\mu$.

The conservation law $\di_\mu \hat I^\mu=0$ has been studied in KBL97.
Therefore the present section is mainly mathematical; it gives the
necessary ingredients for sections 3 and 4.

\noindent (ii) {\it The conserved vector density $\hat I^\mu$. }

Let $g_{\mu\nu}(x^\lam)$ be the metric of the perturbed spacetime
${\cal M}$ and $\Bar g_{\mu\nu}$
be the metric of the
background $\Bar {\cal M}$ both with signature $~-2$.
Once we have chosen a smooth global
mapping such that
each point $ P$ of  ${\cal M}$ is mapped on a point $\Bar{P}$ of  $\Bar
{\cal M}$, we can use the convention that $\Bar{P}$ and $P$  shall
always be given the same coordinates $\Bar {x^\mu} = x^\mu$. This
convention implies that coordinate transformations on    ${\cal M}$
inevitably induce the same coordinate transformations with the same
functions on  $\Bar {\cal M}$.  With this convention, such expressions
as  $g_{\mu\nu} - \Bar g_{\mu\nu}$ become true tensors.  However if the
particular mapping has been left unspecified, we are still free to
change it. The form of the equations for perturbations must inevitably
contain a gauge invariance corresponding to this freedom.

Let $R^\lam_{~\nu\rho\sig}$ and $\Bar
R^\lam_{~\nu\rho\sig}$ be the curvature tensors of ${\cal M}$ and
$\Bar{\cal M}$. These are related as follows:
$$
R^\lam_{~\nu\rho\sig} =
\Bar D_\rho \Delta^\lam_{\nu\sig} - \Bar D_\sig \Delta^\lam_{\nu\rho} +
\Delta^\lam_{\rho\eta} \Delta^\eta_{\nu\sig} - \Delta^\lam_{\sig\eta}
\Delta^\eta_{\nu\rho} + \Bar R^\lam_{~\nu\rho\sig}.
\m (2.2)
$$
Here $\Bar D_\rho$ are covariant derivatives with respect to $\Bar
g_{\mu\nu}$ and $\Delta^\lam_{\mu\nu}$ is the difference between
Christoffel symbols in ${\cal M}$ and $\Bar{\cal M}$:
$$
\Delta^\lam_{\mu\nu} = \Gamma^\lam_{\mu\nu} - \Bar \Gamma^\lam_{\mu\nu}
= \half g^{\lam\rho} \l(\Bar D_\mu g_{\rho\nu} + \Bar D_\nu g_{\rho\mu}
- \Bar D_\rho g_{\mu\nu}\r).
\m(2.3)
$$
Our quadratic Lagrangian density $\lag_G$ for the gravitational field
is here defined by Eq. (1.1) with $\hat k^\mu$ given in Eq. (1.2).
The caret means, as we said before, multiplication by $\sqrt{-g}$,
never by $\sqrt{-\Bar g}$. Thus, if $\hat R = \sqrt{-g} R$,
$\Bar{\hat R}$ will unambiguously mean $\sqrt{-\Bar g}\Bar R$. Notice
that $\hat {\Bar R} = \sqrt{- g}\Bar R \neq \Bar {\hat R} =
\sqrt{-\Bar g}~{\Bar R}$.  The  vector density
$\hat k^\mu$ can also be written in the following form that is often
useful in calculations:
$$
\hat k^\mu = {1\over \sqrt{-g}} \Bar D_\nu\l(-gg^{\mu\nu}\r) = \hat
g^{\mu\rho}\Delta^\sig_{\rho\sig} - \hat g^{\rho\sig}
 \Delta^\mu_{\rho\sig},
\m(2.4)
$$
$\di_\mu \hat k^\mu$ cancels  second order derivatives
of $g_{\mu\nu}$ in $\hat R$.
$\lag$ is the Lagrangian density used by Rosen [19].
$\Bar {\lag}$ is $\lag$ in which $g_{\mu\nu}$ has been replaced by
$\Bar g_{\mu\nu}$. When $g_{\mu\nu} =
\Bar g_{\mu\nu}$, $\lag_G$ is thus
identically zero.  The following formula, deduced
from Eqs. (2.2) and (2.4), shows explicitly how $\lag_G$ is quadratic
in the first order derivatives of $g_{\mu\nu}$ or, equivalently,
quadratic in $\Delta^\mu_{\rho\sig}$:
$$
 \lag_G = {1\over 2\k}
\hat g^{\mu\nu} \l(\Delta^\rho_{\mu\nu} \Delta^\sig_{\rho\sig} -
\Delta^\rho_{\mu\sig} \Delta^\sig_{\rho\nu}\r) -{1\over 2\k}
\hat {~~l^{\mu\nu}}  \Bar R_{\mu\nu}.
\m(2.5)
$$
where $\hat l^{\mu\nu}$ is the perturbed metric density defined in
Eq.  (1.7). If the background is flat and denoted $\Bar {\cal M}_0$ and
if we use Minkowski coordinates $X^\mu$, then $\Bar g_{\mu\nu} =
\eta_{\mu\nu}
= diag(1,-1,-1,-1)$, $~\Bar \Gamma^\lam_{\mu\nu} = 0$ and
$$
\lag_G = - {1\over 2\k}
\l(\hat R + \di_\mu \hat
k^\mu
\r) = {1\over 2\k} \hg^{\mu\nu}\l(
\Gamma^\sig_{\mu\rho} \Gamma^\rho_{\sig\nu} -
\Gamma^\rho_{\mu\nu} \Gamma^\sig_{\rho\sig}\r)
\m(2.6)
$$
which is Einstein's [29] Lagrangian.  $\lag_G$ is thus a
generalization of Einstein's Lagrangian density to perturbations on  a
curved background.

Lie differentials are particularly convenient in describing
infinitesimal displacements in both ${\cal M}$ and $\Bar {\cal M}$; if
the mapping was defined before the displacements it remains defined
after displacements. Let $\Delta x^\mu = \xi^\mu \Delta \lam$ represent an
infinitisimal one-parameter displacement generated by a sufficiently
smooth vector field $\xi^\mu$, the corresponding infinitesimal change
in tensors are given in terms of  Lie derivatives with respect  to this
vector field $\xi^\mu$, $\Delta g_{\mu\nu} = \lix  g_{\mu\nu} \Delta
\lam$,
{\it etc.} The Lie derivatives may be written in terms of  partial
derivatives $\di_\mu$, covariant derivative $\Bar D_\mu$ with respect
to $\Bar g_{\mu\nu}$, or covariant derivative $D_\mu$ with respect to
$g_{\mu\nu}$.  Thus,
$$
\eqalignno
{\Lix  g_{\mu\nu} & =  g_{\mu\lam} \di_\nu \xi^\lam +
g_{\nu\lam} \di_\mu \xi^\lam +
\xi^\lam \di_\lam g_{\mu\nu} &
(2.7a)\cr
 &= g_{\mu\lam} \Bar D_\nu \xi^\lam +
g_{\nu\lam} \Bar D_\mu \xi^\lam +
\xi^\lam \Bar D_\lam g_{\mu\nu} &
(2.7b) \cr  &
= g_{\mu\lam}D_\nu \xi^\lam +
g_{\nu\lam}D_\mu \xi^\lam.&
(2.7c)\cr}
$$
Consider now the Lie derivative $\lix\lag$ of $\lag$ in Eq. (1.1),
not of $\lag_G$.
The Lie derivative of a scalar density like $\lag $ is  the ordinary
divergence  $\di_\mu (\lag\xi^\mu)$.  With the variational principle in
mind we can thus write the following identity
$$
\Lix \lag  = {1\over 2\k} \hat G^{\mu\nu} \Lix  g_{\mu\nu} - {1\over 2\k}
\di_\mu \l(\hat g^{\rho\sig}\Lix
\Gamma^\mu_{\rho\sig} - \hat g^{\mu\rho}\Lix
\Gamma^\sig_{\rho\sig} + \Lix \hat k^\mu \r) = \di_\mu\l(\lag \xi^\mu\r)
\m(2.8)
$$
where Einstein's tensor density $\hat G^{\mu\nu} = \hat R^{\mu\nu}  -
\half g^{\mu\nu}\hat R$ is the variational derivative of $2\k\lag$ with
respect to $g_{\mu\nu}$.  Equation (2.8) is easily converted into a
conservation law by treating the first term after the first equality sign as follows: (a)
replace
$\Lix g_{\mu\nu}$ by the expression (2.7{\it c}),
(b) use the contracted Bianchi
identities $D_\nu G^{\mu\nu} = 0$ and (c) use Einstein's equations
 $\hat G^\mu_\nu = \k \hat T^\mu_\nu$. Identity (2.8) becomes then a
conservation law of this form
$$
\di_\mu \hat i^\mu = 0, ~~~~{\rm with}~~~~\hat i^\mu =\hat
T^\mu_\nu\xi^\nu  - {1\over 2\k}
\l(\hat g^{\rho\sig}\Lix
\Gamma^\mu_{\rho\sig} - \hat g^{\mu\rho}\Lix
\Gamma^\sig_{\rho\sig} + \Lix \hat k^\mu \r) -\lag \xi^\mu.
\m(2.9)
$$
Now comes another exercise which consists in replacing $\Lix g_{\mu\nu}$ and
the D-derivatives of $\Lix g_{\mu\nu}$ that appear in
$\Lix \Gamma^\mu_{\rho\sig}$ - see Eq.
(2.20) below - by  $\Bar D$-derivatives and $\Bar D~\Bar D$-derivatives using Eq.
(2.7{\it b})  this time.  The relation between the two kinds of derivatives, $D$
and $\Bar D$ is best
illustrated on the following simple case
$$
D_\nu \xi^\mu = \di_\nu\xi^\mu + \Gamma^\mu_{\nu\rho}\xi^\rho=\Bar D_\nu
\xi^\mu +
\Del^\mu_{\nu\rho}\xi^\rho.
\m(2.10)
$$
The relations is: (a) write D-derivatives in terms of $\di$-derivatives
and $\Ga$'s  (b) replace the $\di$'s by $\Bar D$'s and $\Ga$'s by
$\Del$'s.  If we operate like that on the terms between parenthesis of $\hat i^\mu$ in Eq.
(2.9), we obtain after a tedious  but quite straightforward calculation the following
result.
$\hat i^\mu$ has a term in $\xi^\mu$, one in $\Bar D_\rho \xi_\sig$ and one that contains
the derivatives of the Lie derivatives of the background metric or of
\footnote {***}{The presence of $\bar z_{\rho\sig}$ comes from replacing second
derivatives using the following
identity
$
\Bar D_{\rho\sig}\xi^\mu = \bar R^\mu_{~\sig\rho\nu}\xi^\nu + 2\Bar
D_{(\rho}\bar z^\mu_{\sig)}
- \Bar D^\mu \bar z_{\rho\sig}$.}
$$
\bar z_{\rho\sig} \equiv \half\Lix \Bg_{\rho\sig} =
\Bar D_{(\rho} \xi_{\sig)}.
\m(2.11)
$$
Here $\xi_\sig = \Bar g_{\sig\mu} \xi^\mu$. Indices will always
been
displaced with the background metric $\Bar g_{\mu\nu}$, never
with $g_{\mu\nu}$.
Thus $\hat i^\mu$ has this form
$$
\hat i^\mu = \l(\hat T^\mu_\nu + {1\over 2\k}
\hat g^{\rho\sig} \Bar R_{\rho\sig}\delta^\mu_\nu +
\hat t^\mu_\nu\r)\xi^\nu +
\hat \sig^{\mu\rho\sig}\Bar D_{\rho}\xi_{\sig}+ \hat e^\mu.
\m(2.12)
$$
The undefined symbols in Eq. (2.12) satisfy the following equalities,
$$
\eqalign
{2\k \hat t^\mu_\nu & =
\hat g^{\rho\sig} \l[\l(\Delta^\lam_{\rho\lam} \Delta^\mu_{\sig\nu}  +
\Delta^\mu_{\rho\sig} \Delta^\lam_{\lam\nu} -
2\Delta^\mu_{\rho\lam} \Delta^\lam_{\sig\nu}\r) -
\delta^\mu_\nu \l(\Delta^\eta_{\rho\sig} \Delta^\lam_{\eta\lam} -
\Delta^\eta_{\rho\lam} \Delta^\lam_{\eta\sig}\r)\r] \cr & +
\hg^{\mu\lam}\l(\Delta^\sig_{\rho\sig} \Delta^\rho_{\lam\nu} -
\Delta^\sig_{\lam\sig} \Delta^\rho_{\rho\nu}\r),}
\m(2.13)
$$
$$
2\k \hat \sig^{\mu\rho\sig} =
(g^{\mu\rho} \Bar g^{\sig\nu} +\Bar g^{\mu\sig} g^{\rho\nu} -
g^{\mu\nu} \Bar g^{\rho\sig})\hat \Del^\lam_{\nu\lam}  -
(g^{\nu\rho} \Bar g^{\sig\lam} +\Bar g^{\nu\sig} g^{\rho\lam} -
g^{\nu\lam} \Bar g^{\rho\sig})  \hat \Del^\mu_{\nu\lam}
\m(2.14)
$$
and
$$
2\k\hat e^\mu = \hg^{\mu\lam} \di_\lam \bar z + \hg^{\rho\sig}\l(
\Bar D^\mu \bar z_{\rho\sig} - 2\Bar D_\rho \bar z^\mu_\sig\r),~~~~~\bar
z=\Bar
g^{\rho\sig}\bar z_{\rho\sig}=\Bar D_\lambda\xi^\lambda.
\m(2.15)
$$

Had we applied Eq. (2.8) to $\Bar {\lag}$  instead of $\lag$ , we
would have written everywhere $\Bar g_{\mu\nu}$ instead of
$g_{\mu\nu}$, from Eq. (2.8) up to Eq. (2.15). We would have found
barred, conserved vector densities $\Bar {\hat i^\mu}$  instead of
${\hat i^\mu}$ that are as follows:
$$
\Bar {\hat i^\mu} = \l(\Bar {\hat T^\mu_\nu}  +
{1\over 2\k} \Bar{\hat R}\delta^\mu_\nu\r)\xi^\nu +
\Bar{ \hat e^\mu}.
\m(2.16)
$$
The simpler form of Eq. (2.16)  compared with Eq. (2.12) comes from
the fact that
$\Bar{\Delta^\mu_{\rho\sig}}=0$ and thus $\Bar {\hat t^\mu_\nu}=\Bar{\hat
\sig^{\mu\rho\sig}}=0$.
Conserved vector densities for $\lag_G = \lag - \Bar {\lag}$
are thus obtained by subtracting Eq. (2.16) from Eq. (2.12). We find
in this way,   a conserved vector density relative to the background
$\hat I^\mu$:
$$
\hat I^\mu = \hat i^\mu - \Bar {\hat i^\mu}= \hat \theta^\mu_\nu \xi^\nu
+ \hat \sig^{\mu\rho\sig}
\Bar D_{\rho}\xi_{\sig} + \hat \eta^\mu,
~~~~~~\di_\mu \hat I^\mu = 0,
\m(2.17)
$$
with
$$
 \hat \theta^\mu_\nu = \hat T^\mu_\nu -
\Bar {\hat T^\mu_\nu} +
{1\over 2\k}
\hat l^{\rho\sig} \Bar R_{\rho\sig} \del^\mu_\nu + \hat t^\mu_\nu,
\m(2.18)
$$
and
$$
{\hat \eta^\mu} =\hat e^\mu - \Bar{\hat e^\mu}={1\over 2\k}\l [
 \hat l^{\mu\lam} \di_\lam \bar z + \hat l^{\rho\sig}\l(
\Bar D^\mu \bar z_{\rho\sig} - 2\Bar D_\rho \bar z^\mu_\sig \r)\r ].
\m(2.19)
$$
For Killing vectors $\bar \xi^\mu$, ${\hat \eta^\mu}=0$ and in Eq.
(2.17), $\hat \sig^{\mu\rho\sig} \Bar D_{(\rho}\bar \xi_{\sig)}$ is
equally zero.  The remaining part of the $\sig$-term  contains the
antisymmetric part $\hat \sig^{\mu[\rho\sig]}$ to which we refereed in
the beginning of this section.  It plays the role of a (relative)
helicity in linearized quantum gravity [30] and, see subsection
3(iii) below, is similar to the helicity in electromagnetism
[31].

\noindent (iii) {\it The superpotential. }

Had we not replaced   $(1/{ \k}) \hat G^\mu_\nu$ by  $\hat T^\mu_\nu$
in Eq. (2.9) the conservation law $\di_\mu\hat i^\mu=0$ would have
remained an identity instead of becoming an equation.
Thus $\hat i^\mu$ {\it must} be equal to the
divergence of an antisymmetric tensor density or ``superpotential''
[32] for any
$g_{\mu\nu}$, $\Bar g_{\mu\nu}$ and $\xi^\mu$, i.e., there
exists a $\hat i^{\mu\nu}=-\hat i^{\nu\mu} $ such that
$~\hat i^\mu=\di_\nu \hat i^{\mu\nu}$.
A superpotential is easily found by replacing in Eq.
(2.9)
$\Lix
\Gamma^\mu_{\rho\sig}$ by its expression in terms of $D_\nu\Lix
g_{\rho\sig}$:
$$
\Lix
\Gamma^\mu_{\rho\sig}= {1\over 2}g^{\mu\nu}\l(D_\rho\Lix
g_{\nu\sig}+D_\sig\Lix g_{\nu\rho}-D_\nu\Lix g_{\rho\sig}\r).
\m (2.20)
$$
The conserved current itself $\hat I^\mu = \hat i^\mu - \Bar
{\hat i^\mu}$
is thus also equal to the divergence of a superpotential  $\hat
I^{\mu\nu}=
\hat i^{\mu\nu} -
\Bar {\hat i^{\mu\nu}}$. This superpotential is
[see [6], on flat backgrounds see  [16]; notice that flatness makes
no difference in Eq. (2.21)]
$$
{\hat I^{\mu\nu}} ={1\over
\k}\big({D^{[\mu}\hat
\xi^{\nu]}}-
\Bar{D^{[\mu}\hat \xi^{\nu]}}\big)+ {1\over \k}\hat \xi^{[\mu}
k^{\nu]}=\hat K^{\mu\nu}
+ {1\over \k}\hat \xi^{[\mu} k^{\nu]},~~~~~~ {\hat I^\mu}=
\di_\nu{\hat I^{\mu\nu}}.
\m(2.21)
$$
In Eq. (2.21) $\hat K^{\mu\nu}$ may be called  the ``relative Komar
superpotential'', relative to the background because
$(1/{\k})D^{[\mu}\hat\xi^{\nu]}$, obtained with the Hilbert Lagrangian,
is known as (half) the Komar [13] superpotential.
$\hat
I^{\mu\nu}$ is linear in $\xi^{\nu}$ and its first derivatives
and, using Eqs. (2.4) and (2.10), can be written as follows
$$
\hat I^{\mu\nu}= {1 \over \k} \hat l^{\lambda[\mu}\Bar D_\lambda\xi^{\nu]}
+ \hat F^{\mu\nu}_{~~~\lambda} \xi^\lambda~~~~{\rm with} ~~~~
\hat F^{\mu\nu}_{~~~\lambda}= {1 \over 2\k}\check{g}_{\lambda\rho} \Bar
D_\sig\l(\hat
g^{\rho[\mu}
\hat g^{\nu]\sig}\r).
\m(2.22)
$$
where $\check{g}_{\mu\nu}$ is the inverse of $\hat {g}^{\mu\nu}$. The
tensor density $\hat F^{\mu\nu}_{~~~\lambda}$ is  Freud's [33]
superpotential on a curved background which has been written in this
form already by Cornish [34] (more on this in section 5).  The great
advantage of a superpotential is, as we said, to replace volume
integrals of complicated vectors by surface integrals of relatively
simple tensors.

\noindent (iv) {\it How does  $ \hat I^{\mu}$ depend on a
divergence in the  Lagrangian density?}

We found so far that N\oe ther's method applied to the perturbed
Lagrangian density, Eq. (1.1) with Eq. (1.2), generates with
every smooth
vector $\xi^\mu$  a conserved vector density $ \hat I^{\mu}$,
Eq. (2.17), the divergence of a superpotential $\hat I^{\mu\nu}$,
Eq. (2.21) or Eq.
(2.22), and both depend on the divergence $\di_\mu \hat k^\mu $. The
contribution of the divergence to our superpotential is apparent in
(2.21).  What is the contribution of the divergence to the different
parts of the conserved vector  density: the energy-momentum
tensor density $\hat \theta^\mu_\nu $ and the helicity tensor
density
$\hat {\sig}^{\mu\rho\sig}$? To find this out let us write
the divergence of the second term of Eq. (2.21) in  a form similar to Eq.
(2.17) with a term
in $\xi^\mu$ and one in $\Bar D_{\rho}\xi_{\sig}$ and
valid for any $\hat k^\mu$:
$$
\di_\nu \l({1\over \k}\xi^{[\mu}\hat k^{\nu]}\r) = {1\over
\k}\Bar{D}_\lam
\l(\delta_\nu^{[\mu}\hat k^{\lam ]}\r)\xi^\nu  +  {1\over \k}\hat
k^{[\rho} \Bar
g^{\mu]\sig}\Bar D_{\rho}\xi_{\sig}.
\m(2.23)
$$
The factors of $\xi^\nu$ and $\Bar D_{\rho}\xi_{\sig}$  represent the
respective contributions  in  $\hat I^\mu$ to
$\hat
\theta^\mu_\nu$ and  $\hat \sig^{\mu\rho\sig}$. There is no
contribution to
$\hat\eta^\mu$. If, in accordance with Eq. (2.1),  we indicate by a
prime the parts of those tensors that
are independent of the divergence we may write, see Eqs. (2.17) and
(2.21),
$$
\hat \theta^\mu_\nu = \hat \theta'^\mu_\nu~ +  {1\over
2\k}\l(\delta^{\mu}_{\nu}
 \Bar D_\lam \hat k^\lam - \Bar D_\nu \hat k^\mu\r),
\m (2.24a)
$$
$$
\hat \sig^{\mu\rho\sig} = \hat \sig'^{\mu\rho\sig} + {1\over 2\k}\l(\hat
k^{\rho} \Bar
g^{\mu\sig}- \hat k^{\mu} \Bar g^{\rho\sig}\r).
\m(2.24b)
$$
A summary of the results obtained in this section
together with comments has already been given in subsection (i).

\vskip .1 in

\beginsection {3. Belinfante's method and Rosenfeld's identities}

(i) {\it Motivations and summary of results.}

In this section we modify the conserved vector densities $\hat I^\mu$
obtained
in Eq. (2.17) using Belinfante's [20] trick. {\it In classical field
theory} on a Minkowski spacetime there is a problem with the canonical
energy-momentum tensor, the equivalent of $\hat \theta^{\mu\nu} = \hat
\theta^\mu_\rho \bar g^{\rho\nu}$ here. The tensor is divergenceless
but not symmetrical and therefore in Minkowski coordinates the
angular momentum tensor is not divergenceless; the total angular
momentum is not conserved. The reason is that it does not take account
of  the spin of the field. It is this situation that Belinfante did
remedy by changing the canonical energy-momentum tensor in such a way
that the total energy momentum would remain unchanged and even the
local density would still remain the same in the appropriate gauge.
Rosenfeld [27] found the  same correction independently and in
arbitrary coordinates. We shall use Rosenfeld's method  in the
next section for a different purpose. The Belinfante correction has
 been applied
to gravity in general relativity on a flat background  in Minkowski
coordinates by Papapetrou [21]. He could then calculate the total
angular momentum at spatial infinity and give a physical meaning to
some of the irreducible coefficients in asymptotic solutions of
Einstein's equations.

Here we first apply  the method to the electromagnetic field on a
curved background. The quantities involved are familiar in
electrodynamics and the results  illustrate well the effect of
Belinfante's modification. We then  apply the same method to the
conserved vector $\hat I^\mu$ of a perturbed gravitational field  on an
arbitrary background. We obtain in this way a new conserved vector
density $\hat {\cal I}^\mu$, see Eq. (3.9), which generates a new
energy tensor density $\hat {\cal T}^\mu_\nu$, Eq.  (3.10) with Eq.
(3.8). There is also a new vector density $\hat {\cal Z}^\mu$ the
analogue, in Eq. (2.17), of the sum $\hat \sig^{\mu(\rho\sig)} \bar
z_{\rho\sig} + \hat \eta^\mu$ which is zero for Killing displacements;
it is defined in Eq. (3.10) and written explicitly in Eq.  (3.25).  We
construct also a new superpotential $\hat {\cal I}^{\mu\nu}$ which is
of great simplicity, see Eqs. (3.18) and (3.19).  Parts of it are
quite familiar in the weak field approximation.  The new $\hat {\cal
I}^{\mu\nu}$ discussed in some detail below in (iv) is remarkably
linear in $\hat l^{\mu\nu}$.  Even more remarkable is that the new
conserved vector  is independent of any divergence that is added to the
Hilbert Lagrangian. The energy-momentum tensor given in Eq.  (3.33)
with Eq.  (3.34) is particularly interesting and its properties are
brought forward by Rosenfeld's identities (3.27) and (3.28).  Identity
(3.28) together with Eq. (3.22) shows clearly that the energy-momentum
tensor density $\hat {\cal T}^{\mu\nu}$ is symmetrical only on
Einstein space backgrounds defined by Eq. (3.31).  Identity (3.27)
shows that on such backgrounds the energy-momentum tensor density is
also divergenceless.

\noindent (ii) { \it Belinfante's correction in electro-magnetism.}

The familiar example of an electromagnetic field in empty space will help to see
what Belinfante's addition does to the canonical
energy tensor. Let $\sqrt{-\bar g}{\cal L}^{\dag} $ be the
Lagrangian density on a flat background for
simplicity. Thus $\Bar R^\lam_{\nu\rho\sig}=0$
 and in arbitrary coordinates:
$$
\sqrt{-\bar g}{\cal{L}}^{\dag}=-{1\over 16\pi}\sqrt{-\bar g}\bar
g^{\mu\rho}\bar
g^{\nu\sig}F_{\mu\nu}F_{\rho\sig},~~~~~~F_{\mu\nu}=\di_\mu A_\nu -
\di_\nu A_\mu.
\m (3.1)
$$
We now repeat on $\sqrt{-\bar g}{\cal{L}}^{\dag}$ the operations performed
on
$\lag$ from Eq. (2.7{\it a}) till Eq. (2.19).
However, to
avoid complications inherent to arbitrary $\xi^\mu$'s, we use only
Killing vectors of
the background for which $\bar z_{\rho\sig}=0$ but we stick to arbitrary
coordinates. Then, in  (daggered) notations similar to
those of last section, we find that if Maxwell's equations hold $\Bar
D_\nu
F^{\mu\nu}=0$, the conserved {\it vector} is of the same form as $ I^{\mu}$
in Eq. (2.17) without a $ \eta^{\mu}$:
$$
 I^{\dag\mu}=  \theta^{\dag\mu}_{~\nu}\bar \xi^\nu +
\sig^{\dag\mu\rho\sig}\di_{[\rho}\bar\xi_{\sig]},
\m(3.2)
$$
Here $ \sig^{\dag\mu\rho\sig}$ is antisymmetric in $\rho\sig$. The
canonical energy-momentum tensor $\theta^{\dag\mu}_{~\nu}$
and the helicity tensor are respectively given by
$$
\theta^{\dag\mu}_{~\nu}= -{1\over 4\pi}\l(F^{\mu\rho}\Bar D_\nu
A_\rho - {1\over 4}F^{\rho\sig}F_{\rho\sig}\delta^\mu_\nu\r), ~~~~~~
\sig^{\dag\mu\rho\sig}= -{1\over 4\pi}F^{\mu[\rho}A^{\sig]}.
\m (3.3)
$$
The Belinfante modification consists in changing $ I^{\dag\mu}$ to
$$
 {\cal I}^{\dag\mu} =  I^{\dag\mu}+ \Bar D_\nu ( S^{\dag\mu\nu\rho}\bar
\xi_
\rho)
\m (3.4)
$$
in  which
$$
S^{\dag\mu\nu\rho}=-S^{\dag\nu\mu\rho}=
\sig^{\dag\rho[\mu\nu]}+
\sig^{\dag\mu[\rho\nu]}-
\sig^{\dag\nu[\rho\mu]}= {1\over 4\pi} F^{\mu\nu}A^{\rho}.
\m (3.5)
$$
The modified current ${\cal I}^{\dag\mu}$ is now of the form
$$
{\cal I}^{\dag\mu}= {\cal T}^{\dag\mu\nu}\bar \xi_\nu,  ~~~~~~
 {\cal T}^{\dag\mu\nu} = {1\over
4\pi}\l(F^{\mu\rho}F_{\rho}^{~\nu}+{1\over 4}\bar g^{\mu\nu}
F^{\rho\sig}F_{\rho\sig}\r).
\m (3.6)
$$
This is the familiar symmetrical, divergenceless,
electro-magnetic field energy-momentum tensor.

\noindent (iii) {\it Belinfante's correction for the conserved
vectors in general relativity.}

We now define, by analogy with Eq. (3.4), a new conserved vector
density $\hat {\cal I}^\mu$ by adding to $\hat I^\mu$,
see Eq. (2.17), a divergence of an anti-symmetric tensor density
constructed with the anti-symmetric part
$\hat\sig^{\mu[\rho\sig]}$ of $\hat\sig^{\mu\rho\sig}$
obtained from Eq. (2.14):
$$
\hat{\cal I}^\mu = \hat I^\mu + \di_\nu\l(\hat{S}^{\mu\nu\rho}\xi_\rho\r)
=
\di_\nu\l(\hat I^{\mu\nu}+\hat{S}^{\mu\nu\rho}\xi_\rho\r) = \di_\nu\hat
 {\cal I}^{\mu\nu}
\m(3.7)
$$
with
$$\hat{S}^{\mu\nu\rho}= -
\hat{S}^{\nu\mu\rho}=\hat\sig^{\rho[\mu\nu]}+
\hat\sig^{\mu[\rho\nu]}-\hat\sig^{\nu[\rho\mu]}.
\m (3.8)
$$

The divergence added to $\hat I^\mu$ is the Belinfante addition in
arbitrary
coordinates. The vector density  $\hat {\cal I}^\mu$ is linear in
$\xi^\mu$, in
$\bar z_{\rho\sig}$,  see Eq. (2.11), and its derivatives $\Bar
D_{\lambda}
\bar z_{\rho\sig}$; $~\hat {\cal I}^\mu$ has
no  term in $\di_{[\rho} \xi_{\sig]}$ anymore. The new conserved
current is thus of the form
$$
\hat {\cal I}^\mu = \hat {\cal T}^\mu_\nu\xi^\nu + \hat {\cal Z}^\mu =
\di_\nu\hat
 {\cal I}^{\mu\nu}
\m(3.9)
$$
with
$$
 \hat {\cal T}^\mu_\nu= \hat \theta^\mu_\nu + \bar D_\rho
\hat S^{\mu\rho}_{~~~\nu},
~~~~~~\hat {\cal Z}^\mu= (\hat {\sig}^{\mu\rho\sig}+
\hat{S}^{\mu\rho\sig})\bar z_{\rho\sig}+ \hat
\eta^\mu
\m(3.10)
$$
while
$$
\hat
 {\cal I}^{\mu\nu}= \hat
  I^{\mu\nu}+\hat {S}^{\mu\nu\rho}\xi_\rho.
\m (3.11)
$$
It can be seen that if $\xi^\mu$ is a Killing vector of the
background
$\bar
\xi^\mu$ for which $\bar z_{\rho\sig}=\hat \eta^\mu=0$, the conserved
vector
is simply
given by
$$
\hat {\cal J}^\mu=\hat {\cal T}^\mu_\nu \bar
\xi^\nu = \di_\nu \hat {\cal J}^{\mu\nu}.
\m (3.12)
$$
where $\hat {\cal J}^\mu= \hat {\cal I}^\mu(\xi=\bar \xi)$ and $\hat {\cal J}^{\mu\nu}=\hat
{\cal I}^{\mu\nu}(\xi=\bar \xi)$.

Now consider for a moment the tensor densities $\hat
\sig^{\mu[\rho\sig]}$ which makes up $\hat {S}^{\mu\nu\rho}$. The
contribution from any {\it any}
$\hat k^\mu$ to the $\sig$-tensors have been singled
out  in
Eq. (2.24{\it b}) from which follows, in particular,  that
$$
\hat
\sig^{\mu[\rho\sig]}=\hat
\sig'^{\mu[\rho\sig]}- {1\over 2\k}\bar g^{\mu[\rho} \hat k^{\sig]}.
\m (3.13)
$$
With this result we can rewrite $\hat {S}^{\mu\nu\rho}$ as a sum of a
$k$-independent
part $\hat {S}'^{\mu\nu\rho}$ and a $k$ contribution and we find that
$$
\hat {S}^{\mu\nu}_{~~\rho}\xi^\rho=\hat {S}'^{\mu\nu}_{~~~\rho}\xi^\rho
-{1\over \k}\xi^{[\mu}\hat
k^{\nu]},~~~~~~~ \hat {S}'^{\mu\nu\rho}= \hat\sig'^{\rho[\mu\nu]}+
\hat\sig'^{\mu[\rho\nu]}-\hat\sig'^{\nu[\rho\mu]}.
\m(3.14)
$$
Notice that $\hat I^{\mu\nu}$ in Eq. (2.21) contains exactly the same $\xi
k$-term
as $\hat {S}^{\mu\nu}_{~~\rho}\xi^\rho$  but with the opposite sign. Thus
the new
superpotential (3.11)
{\it is unchanged} by $\di_\mu\hat k^\mu$ or any other
divergence
added to $\lag_G$ for that matter.  It depends only on the Hilbert
Lagrangian for
the perturbations $\hat R - \Bar {\hat R}$ and the method used
to generate it.
If we take Eqs. (2.21) and (3.14) into account, we may, instead
of Eq. (3.11), write
$$
\hat
 {\cal I}^{\mu\nu}= \hat
  K^{\mu\nu}+\hat {S}'^{\mu\nu\rho}\xi_\rho.
\m(3.15)
$$
The divergence of this \sss  is equal to the new conserved vector $\hat
{\cal I}^\mu$. Consequently $\hat {\cal T}^\mu_\nu$ and $\hat {\cal
Z}^\mu$, are independent of any divergence added to the Lagrangian,
contrary to the canonical energy tensor obtained by N\oe ther's method
alone. A direct calculation, {\it without using \ssss } gives of
course the same results. Calculations are only more cumbersome.

We shall now work out explicitly the formulas for $\hat {\cal
I}^{\mu\nu}$ and for $\hat {\cal Z}^\mu$. The explicit form of $\hat
{\cal T}^\mu_\nu$ is obtained below with Rosenfeld's identities.

\noindent (iv) {\it The explicit form of the superpotential $\hat
{\cal I}^{\mu\nu}$.}

The tensor $\hat\sig'^{\mu[\rho\sig]}$ is obtained
from Eq. (3.13) in
terms of
$\Delta$'s using  Eq. (2.14) for
$\hat\sig^{\mu[\rho\sig]}$ and Eq. (2.4) for $\hat k^\mu$. The sum of
the $\sig'$'s that make up $\kappa \hat {S}'^{\mu\nu\sig}$ in Eq. (3.14) is
found to give
$$
\k\hat {S}'^{\mu\nu\sig}= \l(\Delta^\rho_
{\rho\lambda}\hat g^{\sig[\mu} - \Delta ^\sig_
{\rho\lambda}\hat g^{\rho[\mu}\r)\bar g^{\nu]\lambda} +\l(\hat
g^{\sig\rho}\bar g^{\lambda[\mu}
-  \bar g^{\sig\rho}\hat g^{\lambda[\mu}\r) \Delta ^{\nu]}_
{\rho\lambda}+
\hat g^{\rho\lambda} \Delta ^{[\mu}_ {\rho\lambda}\bar g^{\nu]\sig}.
\m(3.16)
$$
On the other hand $\kappa \hat K^{\mu\nu}$ defined in Eq. (2.21) can be
written with the help of Eq. (2.10) in this form
$$
\k\hat K^{\mu\nu}= \hat l^{\rho[\mu}\Bar D_\rho\xi^{\nu]} + \hat
g^{\rho[\mu}
\Delta ^{\nu]}_
{\rho\sig}\xi^\sig.
\m(3.17)
$$
The sum of Eq. (3.16) times $\xi_\sig$ and Eq.  (3.17) gives the new
superpotential in terms of
$\Delta$'s. However, $\hat
{\cal I}^{\mu\nu}$ has a much nicer form in terms of $\Bar D$
derivatives of the metric $g_{\mu\nu}$ or rather in terms of $\hat l^{\mu\nu}$. With Eq.
(2.3) we can write $\hat {\cal I}^{\mu\nu}$ in the following form
$$
\hat
{\cal I}^{\mu\nu}= {1 \over \k} \hat l^{\rho[\mu}\Bar D_\rho\xi^{\nu]}
+ \hat {\cal
P}^{\mu\nu}{_\lambda} \xi^\lambda
\m(3.18a)
$$
where the $\cal P$-tensor plays now the role of the $F$ tensor in (2.22):
$$
 \hat {\cal P}^{\mu\nu\rho}=
-{\cal P}^{\nu\mu\rho}= {1 \over 2\k} \Bar D_\sig\l(\bar g^{\rho\mu} \hat
l^{\nu\sig}-\bar g^{\rho\nu} \hat
l^{\mu\sig}-  \bar g^{\sig\mu} \hat
l^{\nu\rho}+\bar g^{\sig\nu} \hat
l^{\mu\rho}\r).
\m(3.18b)
$$
Another telling and useful form of the superpotential is
$$
\hat
{\cal I}^{\mu\nu}= {1 \over \k} \l(\hat l^{\rho[\mu}\Bar D_\rho\xi^{\nu]}+ \xi^{[\mu}\Bar D_\sig \hat
l^{\nu]\sig}-\bar D^{[\mu}\hat l^{\nu ]}_\sig \xi^\sig \r).
\m (3.19)
$$
The  superpotential (3.19) {\it linear} in $\hat l^{\mu\nu }$ and its
derivatives is (not surprisingly) reminiscent of various familiar
expressions used in the literature  and which are based on linear
approximations or expansions to higher orders:

\noindent (a) On a flat background:

\noindent In Minkowski coordinates $X^\mu$, $\bar
g_{\mu\nu} =
 \eta_{\mu\nu}= diag(1,-1,-1,-1)$. There are ten Killing vectors in the
background. The four components of the  Killing vectors of
translations can be taken equal to $\bar \xi^\mu =
\delta^\mu_\alpha$ with $\alpha = (0,1,2,3)$ and  $\Bar D_\nu\bar
\xi^\mu=0$; thus the four corresponding superpotentials are
$$
 \hat {\cal P}^{\mu\nu\alpha}=
{1\over 2\k}\di_\sig\l(\eta^{\mu\alpha}\hat
g^{\nu\sig}+\eta^{\sig\nu}\hat g^{\mu\alpha}-\eta^{\alpha\nu}\hat
g^{\mu\sig}-\eta^{\sig\mu}\hat g^{\nu\alpha}\r).
\m(3.20)
$$
This anti-symmetric  tensor density $\hat {\cal P}^{\mu\nu\alpha}$ is
sometimes quoted as the Papapetrou superpotential [21]. Considered as a
linearized approximation $\hat {\cal P}^{\mu\nu\alpha}$ is the same
quantity as Weinberg's [35] $\hat Q^{\mu\nu\alpha}$ and  Misner, Thorne
and Wheeler's [36] $\partial_\beta \hat H^{\mu\alpha\nu\beta}$.
Equation (3.20) is identical with the linearized approximations of
Freud's superpotential, Eq. (2.22), or of the Landau and Lifshitz
superpotential written in arbitrary coordinates like in Cornish [34].

With the three
spatial components for the Killing vectors of rotation which in $X^\mu$
coordinates are given by
$\bar
\xi^\mu= (\delta^\mu_\alpha\eta_{\beta\gamma}-
\delta^\mu_\beta\eta_{\alpha\gamma})
X^\gamma$, $\hat {\cal I}^{\mu\nu}$
becomes {\it the} Papapetrou superpotential for angular momentum
as given in his original paper.

\noindent (b) On  Einstein space backgrounds:

\noindent The linear approximation of  Eq. (3.19) in arbitrary
coordinates   with  Killing vectors $\bxi^\mu$ is equal
to the Abbott and Deser [37] superpotential
worked out on an Einstein space background [Eq. (3.31) below].
To obtain their full non-linear
expression replace $\hat l^{\mu\nu}$ by $-\sqrt
{-\bar g}H^{\mu\nu}$ where
$H^{\mu\nu}=\bar g^{\mu\rho}\bar g^{\nu\sig}H_{\rho\sig}$ with
$H_{\rho\sig}$  defined in Eq. (A.13).

\noindent (c) It is important to note that the correction to
$\hat I^{\mu\nu}$ namely
$\hat {S}^{\mu\nu\rho}\xi_\rho$ is {\it homogeneous of order
two} in $\hat
l^{\mu\nu}$ and its derivatives. This is easily seen with Eq. (2.14)
because
$$
2\k \hat \sig^{\mu[\rho\sig]} =
\l(\hat l^{\mu[\rho} \Bar g^{\sig]\nu} -\Bar g^{\mu[\rho} \hat
l^{\sig]\nu}\r)
\Del^\lam_{\nu\lam}  - \l(\hat l^{\nu[\rho} \Bar g^{\sig]\lam} -\Bar
g^{\nu[\rho}
\hat l^{\sig]\lam}\r) \Del^\mu_{\nu\lam}.
\m(3.21)
$$
Thus $\hat I^{\mu\nu}$ and  $\hat {\cal I}^{\mu\nu}$ are equal in the
linear approximation as well on arbitrary backgrounds. From this follows
[see KBL97] that $\hat {\cal I}^{\mu\nu}$ provides the correct energy
and linear momentum at spatial infinity. It gives also correctly the
Bondi [15] - Sachs [38] energy and linear momentum at null infinity
(see [39]). Details and proofs are
given in appendix where we show also that the Abbott and Deser
\sss does not give the Bondi-Sachs linear momentum.

\noindent (v) {\it The new  $\hat {\cal Z}^\mu$ vector density.}

This vector density plays a role in conservation laws that are not
associated with Killing vectors as in the examples mentioned in the
introduction and in the cosmological applications of section 4.  The
vector density has also an important task in Rosenfeld's identities
(see below). $\hat {\cal Z}^\mu$ is defined in Eq. (3.10).  It must be
noted in Eq. (3.9) that the factor of $\Bar D_\rho \xi_\sig$ {\it is}
indeed symmetrical in $\rho\sig$: This follows from the definition
(3.8) of $\hat S^{\mu\sig\rho}$. We need a new symbol for that factor
which plays a role later on; let us set
$$
*\hat
S^{\mu\rho\sig}=*\hat
S^{\mu\sig\rho}= \hat \sig^{\mu\sig\rho}+\hat
S^{\mu\sig\rho}=\hat\sig^{(\mu\rho)\sig}+
\hat\sig^{(\mu\sig)\rho}-\hat\sig^{(\rho\sig)\mu}=
\hat\sig'^{(\mu\rho)\sig}+
\hat\sig'^{(\mu\sig)\rho}-\hat\sig'^{(\rho\sig)\mu}.
\m(3.22)
$$
Thus
$$
\hat {\cal Z}^\mu =*\hat
S^{\mu\rho\sig}\bar z_{\rho\sig} + \hat \eta^\mu
\m(3.23)
$$
with $\hat \eta^\mu$ defined by Eq. (2.19).
An interesting form of $*\hat
 S^{\mu\rho\sig}$
comes from using Eq. (2.14) for $\hat \sig^{(\mu\rho)\sig}$:
$$
*\hat S^{\mu\rho\sig}
={1\over 2\k}\Bar
D_\nu\l(2\hat
l^{\mu(\rho}\bar g^{\sig)\nu} - \bar g^{\mu\nu}\hat l^{\rho\sig}-\hat
l^{\mu\nu}\bar
g^{\rho\sig}\r).
\m(3.24)
$$
Plugging Eqs. (3.24) and (2.19) into  $\hat{\cal Z}^\mu$ gives the
following form ``anti-symmetric''  in
$ \bar z \leftrightarrow l$
$$
2\k \hat
 {\cal Z}^\mu = 2\l(\bar z^{\rho\sig}\Bar D_\rho \hat l^\mu_\sig -
\hat l^{\rho\sig}\Bar D_\rho \bar z^\mu_\sig\r) - \l(\bar z_{\rho\sig}\Bar
D^\mu \hat
l^{\rho\sig} -
\hat l^{\rho\sig}\Bar D^\mu \bar z_{\rho\sig}\r) + \l(\hat l^{\mu\nu}\Bar
D_\nu \bar z -
\bar z\Bar D_\nu \hat l^{\mu\nu}\r).
\m(3.25)
$$

\noindent (v) {\it The Rosenfeld identities.}

The conservation law $\di_\mu \hat {\cal I}^\mu=0$ which holds for any
smooth
vector $\xi^\mu$
contains derivatives of $\xi^\mu$ of an order as high as 3. Thus with
the help of Eq. (3.9),
$\di_\mu \hat {\cal I}^\mu=0$ can be written in the form
$$
\di_\mu \hat {\cal I}^\mu = \di_\mu(\hat {\cal T}^\mu_\nu\xi^\nu) +
\di_\mu\hat {\cal Z}^\mu=
\hat
\beta_\nu \xi^\nu+
\hat \beta^\mu_{~\nu} \Bar D_\mu \xi^\nu  +
 \hat \beta^{\rho\sig}_{~~~\nu} \Bar D_{(\rho\sig)} \xi^\nu +
 \hat \beta^{\mu\rho\sig}_{~~~~\nu} \Bar D_{(\mu\rho\sig)} \xi^\nu = 0.
\m(3.26)
$$
This identity holds for arbitrary smooth $\xi$'s. Therefore all the
properly symmetrized $\beta$'s must be
identically zero, $\hat
\beta_\nu =
\hat \beta^\mu_{~\nu} =
 \hat \beta^{(\rho\sig)}_{~~~\nu} =
 \hat \beta^{(\mu\rho\sig)}_{~~~~\nu}= 0$. This is the way  in
which Rosenfeld [27] obtained a set of identities and found the
Belinfante
correction.
$\hat {\cal T}^\mu_\nu$ appears obviously in $\hat \beta^\mu_{~\nu}=0$ and
$\Bar D_\mu \hat {\cal T}^\mu_\nu$ in $\hat \beta_\nu=0$.
The remaining identities are independent of the energy tensor. In
Rosenfeld's work, which is in arbitrary coordinates but on a
flat background, $\hat \beta^\mu_{~\nu}=0$ was the most interesting
relation. It related the canonical energy momentum generated by
N\oe ther's method to the symmetric and divergenceless energy-momentum
needed in Einstein's equations. The divergencelessness was shown by
$\hat \beta_\nu =0$.

Here, however, the background is not flat and Rosenfeld's identities
give different and interesting results.
It is obvious that the calculations of the
$\beta$'s asks for a lot of rearrangements of factors that come
exclusively from
$\di_\mu\hat {\cal Z}^\mu$. This is somewhat tedious but really
straightforward. The resulting identities have in the end a nice form:
$$
\hat \beta_\nu = \Bar D_\mu \hat
{\cal T}^\mu_\nu - {1\over 2\k} \hat l^{\rho\sig}\Bar D_\mu \Bar
R_{\rho\sig}=0,
\m (3.27)
$$
$$
\hat \beta^{\mu\nu}= \hat \beta^\mu_{~\rho}\bar g^{\rho\nu}= \hat {\cal
T}^{\mu\nu}+ \Bar
D_\rho(*\hat S^{\rho\mu\nu}) - {1\over \k}\hat l^{\rho\mu}\bar R^\nu_\rho=0,
\m(3.28)
$$
$$
\hat \beta^{(\rho\sig)\nu}= *\hat S^{(\rho\sig)\nu} +
\Bar D_\mu \hat \beta^{\mu\rho\sig\nu}= 0,
\m (3.29)
$$
$$
\hat \beta^{(\mu\rho\sig)\nu}= 0 \qquad{\rm with}\qquad
\hat \beta^{\mu\rho\sig\nu} =
\hat
\beta^{\mu\sig\rho\nu}= {1\over 2\k} \l(\hat l^{\mu(\rho}\bar
g^{\sig)\nu} - \hat
l^{\rho\sig}\bar g^{\mu\nu}\r).
\m (3.30)
$$
If we remember that $*\hat S^{\rho\mu\nu}$ is symmetrical
in $\mu\nu$, see Eq. (3.24), we see from
Eq. (3.28) that $\hat
{\cal T}^{\mu\nu}=\hat
{\cal T}^{\nu\mu}$ if and only if
$$
\hat  l^{\rho[\mu}\bar R^{\nu]}_\rho=0   ~~~ \rightarrow
~~~ \Bar R_{\mu\nu} = - \Bar \Lambda \bar g_{\mu\nu}
\m (3.31)
$$
where $\Bar \Lambda$ is necessarily a constant. Thus the
energy-momentum tensor is symmetrical on backgrounds that are {\it
Einstein spaces} in the sense of A.Z. Petrov [24]. de Sitter and
Einstein's cosmological {\it spacetimes} belong to this class but in
general Friedmann-Robertson-Walker spacetimes do not.  This is why our
formalism with arbitrary $\xi^\mu$'s is precisely good in relativistic
cosmology and why $\hat {\cal Z}^\mu$ may be important in that case.

When the background is an Einstein space, Eq. (3.27) shows that $\hat
{\cal T}^{\mu\nu}$ is also divergenceless. Identity (3.28)  provides an
interesting form
for $\hat {\cal T}^{\mu\nu}$;
using Eq. (3.24) in Eq. (3.28) we can write $\hat
{\cal T}^{\mu\nu}$ like this:
$$
\hat
{\cal T}^{\mu\nu}= -\Bar D_\rho \hat {\cal P}^{\rho\nu\mu} +
{3\over 2\k}\hat l^{\rho[\mu}\bar R^{\nu]}_\rho +
{1\over 2\k}[\hat l^{\rho(\mu}\bar R^{\nu)}_\rho -
\hat l^{\rho\sig}\bar R^{\mu~~\nu}_{~\rho\sig}].
\m (3.32)
$$
On a flat background we  see, looking at Eq. (3.20), that $\hat {\cal
T}^{\mu\nu}$ is the second order derivative of a tensor which was also
obtained by Papapetrou [21]. Thus Eq. (3.32) is the generalization of his
equation to curved backgrounds.

The new energy-momentum tensor can be calculated from Eq. (3.10)
with the use of Eqs. (2.13), (2.14), (2.18) and (3.8). It contains three
types of terms, very similar to those of $\hat \theta^{\mu\nu}$: a
symmetric matter energy momentum of the perturbations, a symmetric
field energy-momentum tensor $\hat \tau^{\mu\nu}=\hat \tau^{\nu\mu}$
and two non-derivative couplings  to the Ricci
tensor of the background, the second of which being antisymmetrical:
$$
\hat {\cal T}^{\mu\nu} = (\hat T^{(\mu}_\rho\Bar g^{\nu)\rho} -\Bar {\hat
T^{\mu\nu}})+\hat \tau^{\mu\nu} +{1\over 2\k}\hat
l^{\rho\sig}
\Bar R_{\rho\sig} \Bar g^{\mu\nu} + {1\over \k} {\hat l}^{\lam[\mu} \bar
R^{\nu]}_\lam.
 \m(3.33)
$$
The field energy-momentum tensor density is the
following terrifying homogeneous quadra-tic
form in $\hat
l^{\mu\nu}$, their first and second order derivatives:
$$
\eqalign
{\k \hat \tau^{\mu\nu} & =
\half \l(\hat l^{\mu\nu}\Bar g^{\rho\sig} -
\Bar g^{\mu\nu}\hat l^{\rho\sig}\r)
\Bar D_\sig \Delta^\lambda_{\rho\lambda} +
  \l(\hat l^{\rho\sig} \Bar g^{\lambda(\mu} -
\Bar g^{\rho\sig}\hat l^{\lambda(\mu}\r)
\Bar D_\sig \Delta^{\nu)}_{\lambda\rho} \cr &
+
\Bar g^{\rho\sig}\l(\half\hat g^{\mu\nu}\Delta^\lambda_{\rho\lambda}
\Delta^\eta_{\sig\eta}+
\hat g^{\lambda\eta}\Delta^{(\mu}_{\lambda\rho}\Delta^{\nu)}_{\eta\sig} +
\Delta^{\lambda}_{\sig\eta}\Delta^{(\mu}_{\lambda\rho}\hat g^{\nu)\eta} -
2\Delta^{\lambda}_{\sig\lambda}\Delta^{(\mu}_{\eta\rho}\hat g^{\nu)\eta}\r)
\cr & +
\hat g^{\lambda\eta}
\l [\half
\Bar
g^{\mu\nu}\Delta^\sig_{\rho\lambda}\Delta^\rho_{\sig\eta}+
\l(\Delta^{\sig}_{\rho\sig}\Delta^{(\mu}_{\lambda\eta}
-
\Delta^{\sig}_{\lambda\eta}\Delta^{(\mu}_{\rho\sig} -
\Delta^{\sig}_{\lambda\rho}\Delta^{(\mu}_{\eta\sig}\r)\Bar g^{\nu)\rho}\r ].}
\m(3.34)
$$
On Ricci flat backgrounds $\hat {\cal T}^{\mu\nu}$, Eq. (3.33), reduces
to the expression found by Grishchuk {\it et al} [40].

\noindent (vii) {\it Second derivatives in the  energy tensor?}

Second derivatives of $g_{\mu\nu}$ appear in the field energy tensor.
This needs some comments. The canonical field energy $\hat t^\mu_\nu$,
see Eq. (2.13), is quadratic in first order derivatives.  Thus that
tensor density depends certainly on initial conditions only. This is
the normal behavior of a conserved quantity.

Notice that the volume integral of the new conserved vector density
$\hat {\cal I}^\mu $ is equal to a surface integral of  $\hat {\cal
I}^{\mu\nu}$ in which there are no more than  first order derivatives.
This is also a suitable result.

Consider now the local quantities $\hat {\cal I}^\mu$ themselves.
Suppose that initial conditions are defined on a hypersurface at a
given time coordinate $x^0=0$. Initial conditions are the metric
components and their time derivatives (modulo Einstein's constraints
and gauge freedom). The coordinate density is equal to  $\hat {\cal
I}^0= \hat I^0 + \di_k (\hat S^{0k\sig}\xi_\sig)$ with $k=1,~2,~3$.
Only spatial derivatives of S appear in $\hat {\cal I}^0$ because $S$  is
anti-symmetric in the first two indices.  Thus since $\hat I^0$ and
$\hat S^{0k\sig}\xi_\sig$ contain only first order {\it time} derivatives,
$\hat {\cal I}^0$ itself contains only first order {\it time}
derivatives and therefore even the energy-momentum density contains no
more than first order time derivatives. The Belinfante correction adds
only second order spatial derivatives. It is sometimes required, like in KBL97
that the energy-momentum tensor should contain no more than first order
derivatives. This requirement may be unnecessarily restrictive.

\noindent (viii) {\it Conservation laws directly from Einstein's equations?}

There is no  difficulty to rebuild Einstein's equations
$\hat G^\mu_\nu=\k \hat
T^\mu_\nu$ from the conservation laws
$\di_\nu\hat{\cal I}^{\mu\nu}=\hat{\cal I}^\mu$. To do this we rewrite
$\hat{\cal I}^{\mu\nu}$ given by
(3.19) as follows
$$
\k \hat{\cal I}^{\mu\nu}= -\bar D^{[\mu}\hat l^{\nu]}_\rho \xi^\rho
+\xi^{[\mu} \hat{\cal G}^{\nu]}+\hat
l^{\rho[\mu}\bar D_\rho \xi^{\nu]}
\m (3.35)
$$
where $\hat{\cal G}^{\nu}=\bar D_\rho \hat l^{\rho\nu}$.
It is useful to keep track of $\hat{\cal G}^{\nu}$ because  $\hat{\cal
G}^{\nu}=0$ is the familiar (generalization of the) well known  De Donder
gauge condition. Thus, if we
take the divergence of Eq. (3.35) and
remember that
$\di_\nu\hat{\cal I}^{\mu\nu}=\hat{\cal I}^\mu$ we find, by replacing
$\hat{\cal I}^\mu$ with $\hat {\cal
T}^\mu_\nu\xi^\nu+\hat {\cal Z}^\mu$ and $\hat {\cal
T}^{\mu\nu}$ by its expression given in Eq. (3.33) that
$$
\eqalign
{ 2\k \di_\nu \hat{\cal I}^{\mu\nu} & =
\l(\bar D_\rho \bar D^\rho \hat l^{\mu\nu}-
2\bar D^{(\mu}\hat{\cal G}^{\nu)}+
\bar D_\rho\hat{\cal G}^{\rho} \bar g^{\mu\nu}+
2\hat l^{\rho[\mu}\bar R^{\nu]}_\rho -
2\hat l^{\rho\sig}\bar R^{\mu~~\nu}_{~\rho\sig} \r )\xi_\nu +
2\k \hat {\cal Z}^\mu \cr & =
2\k\hat{\cal I}^\mu =\l[2\k \l(\hat  T^{(\mu}_\rho \bar
g^{\nu)\rho}- \Bar {\hat T^{\mu\nu}} + \hat \tau^{\mu\nu}\r)
 +2\hat l^{\rho[\mu}\bar R^{\nu]}_\rho + \hat l^{\rho\sig}\bar
R_{\rho\sig}\bar g^{\mu\nu}\r]\xi_\nu +2\k\hat
 {\cal Z}^\mu.}
\m (3.36)
$$
We can remove $2\k\hat
{\cal Z}^\mu$ and $2\hat l^{\rho[\mu}\bar R^{\nu]}_\rho\xi_\nu$ from both
sides of Eq. (3.36). The remaining
homogeneous linear expression in $\xi_\nu$  is true for any $\xi_\nu$. The
factors of $\xi_\nu$  on both sides
of the equality must thus be equal. We are left with a set of equations
that are of course
Einstein's equations in which the left hand side contains
all terms that are linear in $\hat l^{\mu\nu}$:
$$
\bar D_\rho \bar D^\rho \hat l^{\mu\nu}-2\bar D^{(\mu}\hat{\cal
G}^{\nu)}+\bar D_\rho\hat{\cal G}^{\rho} \bar g^{\mu\nu}-\hat
l^{\rho\sig}\bar R_{\rho\sig}\bar g^{\mu\nu}  -2\hat
l^{\rho\sig}\bar R^{\mu~~\nu}_{~\rho\sig}=2\k \l(\hat  T^{(\mu}_\rho \bar
g^{\nu)\rho}- \Bar {\hat T^{\mu\nu}} +
\hat \tau^{\mu\nu}\r).
\m (3.37)
$$
On a flat background, in Minkowski coordinates these are Einstein's
equations as they were written down by Papapetrou [21].  Equations
(3.37) have also been given in this form by Grishchuk {\it et al} [40]
on a Ricci flat background $\bar R_{\rho\sig}=0$.

The linearized
approximation on a flat background with the De Donder gauge
condition is readily  recognized as the
gravitational wave equations written in arbitrary
coordinates:
$$
\bar D_\rho \bar D^\rho (\sqrt{-g}g^{\mu\nu})=2\k \hat  T^{\mu\nu}.
\m (3.38)
$$
Equation (3.37) is an interesting form of Einstein's equations from which we
could have  constructed our conservation laws $\di_\nu\hat{\cal
I}^{\mu\nu}=\hat{\cal I}^\mu$.  But who has thought that by adding
$2\k\hat {\cal Z}^\mu+2\hat l^{\rho[\mu}\bar R^{\nu]}_\rho\xi_\nu$  on
both sides of Eq.  (3.37) we would get
$\di_\nu\hat{\cal I}^{\mu\nu}=\hat{\cal I}^\mu$?

 \vskip .1 in

\nnn {\bf 4. Conformal Killing vectors, conservation laws and
integral constraints}

\nnn {\bf in cosmology}

\nnn {\it (i) Motivations and summary of results.}

Here we illustrate the theory developed in the previous sections with
some applications in theoretical cosmology. Conservation laws have been
used previously in cosmology (see introduction) and the following
examples give potentially useful new formulas.

We start by considering \frw \bbbb with their 15 conformal Killing
vectors. We take the metric of the \bbbb in the form given by Eq. (4.1) and
with Eq. (4.2) in terms of a conformal time $\eta$. In these coordinates the
\ckkkk satisfy the  10 equations given in Eq. (4.6). A set of 15 linearly
independent solutions of those equations is given in Eqs. (4.7) and  (4.8)
for spacetimes with flat spacelike sections $\eta=0$ $(k=0)$ and in  Eqs.
(4.7) and (4.10) for curved spacelike sections $(k=\pm 1)$.  Conserved
quantities and integral constraints at a given time $\eta$ for
perturbations of the \bbb  will be obtained from the time components of
the conserved vector densities $\hat{\cal I}^0$ and the corresponding
\sss components $\hat{\cal I}^{0l}$ $(l=1,2,3)$. These components can
be calculated for the 15 \ckkkk and the general formulas for that are
given by Eqs. (4.18{\it a}) or (4.18{\it b}) and (4.19).
Equation (4.18{\it a}) provide an
expression for perturbations that may even be large. The rest of this
section deals mainly with applications of these formulas.

The 15 integrands $*{\cal I}$ and $*{\cal I}^l$ defined in Eq. (4.20)
are calculated explicitly. Notice  that Eqs. (4.18) and (4.19) contain
components of the \ckkkk and their spatial derivatives. They do not
contain the time derivatives of these components. We may thus use
$*{\cal I}$'s and $*{\cal I}^l$'s obtained from linear combinations of
\ckkkk with time dependent coefficients. The reason for doing this is
that we find in this way 15 relatively simple integrands.  They are
 written in Eqs. (4.24) for $k=0$ and in Eqs. (${\bf \~{\rm{4.24}}}$)
for $k=\pm 1$.
Notice that linear combinations of \ckkkk are not \ckkkk anymore.
Nevertheless with those non \ckkkk we have obtained interesting
``conservation laws".  We  show, for instance, that Traschen's integral
constraint vectors [1] are equal to time dependent linear combinations of
conformal Killing vectors.  Formulae are explicitly given in Eq. (4.25)
for $k=0$ and Eq. (4.26) for $k=\pm 1$.  We also show that if we equate to
zero the uniform Hubble expansion rate ${\cal Q}$ [26] defined in Eq.
(4.17), 14 of the 15 integral constraints take the particularly simple
form shown in Eq. (1.10). The lone exception is associated with time
translations $(k=\pm 1)$ or time accelerations $(k=0)$.  These
exceptions are precisely those conservation laws that interested Uzan
et al [8] (see also [7]) in which an unexpected field contribution in
the conservation law of ``energy'' is inevitable. We also give a non
trivial illustration of a spacetime that is asymptotically
Schwarzschild-de Sitter with $k=0$. In this case we obtain 13 global
integral constraints with no boundary contributions and two non-zero
constants of motion for perturbations that may be large near the source
but weak at infinity.  The results, which may be new, are given in Eq.
(4.34).  For perturbations that are small everywhere, the formulas are
given by Eq. (4.35).

\nnn (ii) {\it \frwwww and their conformal Killing vectors.}

We write the \bbb metric $d\bar s^2$ in dimensionless coordinates
$\xm= (x^0=\eta,~ x^k)$ with $ k,l,m=1,2,3$ for
which the symmetrical role of $x^k$ is  apparent:
$$
d\bar s^2= \bar g_{\mu\nu}dx^\mu dx^\nu=
a^2(d\eta^2-\f.kl dx^kdx^l)= a^2 e_{\mu\nu}dx^\mu dx^\nu,
\m (4.1)
$$
$a(\eta)$ is the scale factor and $\f.kl$, $f^{kl}$ and
$f=det(\f.kl)$ are respectively given by
$$
\f.kl = \delta_{kl} + k {{\delta_{km}x^m \delta_{ln}x^n}\over 1-kr^2  },
 ~~~ f^{kl}=\delta^{kl}-kx^k x^l,~~~
f={1\over 1-kr^2},
\m(4.2)
$$
$k=0$ or $\pm 1$ and $r^2=\delta_{kl}x^k x^l$.
The non-zero Christoffel symbols of the metric (4.1) are
$$
\bar \Gamma^0_{00}=\dot a,~~~\bar \Gamma^0_{kl}=\dot a \f.kl,~~~
\bar \Gamma^m_{0l}=\dot a \delta^m_l,~~~
\bar \Gamma^m_{kl}=kx^m \f.kl,
\m (4.3)
$$
$\dot a$  is the dimensionless conformal Hubble ``constant"
$$
\dot a = {1\over a}{da\over d\eta}.
\m (4.4)
$$
In these notations the non zero components of the
Einstein tensor are respectively
$$
\Bar{G}^0_0={3\over a^2}(k+\dot a^2)=\k\Bar{T}^0_0,~~~
\Bar{G}^m_l = {1\over a^2}(k+\dot a^2+2\di_0 \dot a)\delta^m_l=\k\Bar
{T}^m_l,
\m (4.5)
$$
The \ckkkk are the 15 linearly independent solutions
$\xi^\mu=(\xi^0,\xi^k)$ of Eq. (1.8). These equations are
{\it independent} of the conformal factor and can be written
in 3-dimensional notations as follows:
$$
\di_0{\xi^0}={1\over 3}\nabla_k\xi^k,\qquad  \di_0 {\xi^k}
=\nabla^k \xi^0,\qquad
 \nabla^{(k} \xi^{l)} = f^{kl}\di_0{\xi^0}.
\m (4.6)
$$
where $\nabla_k$ is a 3-covariant derivative for the $\f.kl$ metric,
$\nabla^k=f^{kl}\nabla_l$, and the first equation equals one third of
the trace of the last one. We found the solutions of Eq. (4.6)  as
follows. (a) The group of conformal transformations in Minkowski
coordinates $X^\mu$ is explicitly given in [25]; infinitesimal
transformations provide the \ckkkk of the flat \spttt in $X^\mu$
coordinates. (b) The metric $e_{\mu\nu}$ is conformal to the Minkowski
metric $\eta_{\mu\nu}$, $~$ i.e.  in appropriate coordinates
$e_{\mu\nu}=\Omega^2 \eta_{\mu\nu}$ globally and the \ckkk components
are the same since they do not dependent on $\Omega$; we had thus only
to take the components of $\xi^\mu$ in Minkowski coordinates and
transform them into our  coordinates $x^\mu$; this is easily calculated
from the explicit global coordinate transformations given in [41]. (c) One
can in the end verify  that the results satisfy indeed Eq.  (4.6). Here
are the results.

There are 7
\ckkkk which can be written in compact form for every value of
$k$; these are the \ckkkk of time accelerations ({\bf t}), space
translations (${\bf s}_a,~ a=1,2,3$)
and space rotations (${\bf r}_a$):
$$
t^\mu=\delta^\mu_0,~~~s^\mu_a=\delta^\mu_a \sqrt{1-kr^2},~~~
r^\mu_a=\delta ^{\mu k}\epsilon_{kal}x^l.
\m (4.7)
$$
The other 8 \ckkkk are somewhat different for $k=0$ and for $k=\pm 1$.

For $k=0$, the Lorentz boosts (${\bf l}_a$), dilatation (${\bf d}$), time
acceleration (${\bf a}_0$) and space accelerations (${\bf a}_a$)
(the last two have been
studied in [25]) are respectively given by
$$
\eqalign
{ &{\bf l}_a=(l^0_a=x^a,~{l^k_a}=\eta \delta^k_a),~~
{\bf d}=(d^0=\eta,~ d^k=x^k),\cr &
{\bf a}_0=(a^0_0=\eta^2+r^2, ~ a^k_0 =2\eta x^k), ~~
{\bf a}_a=(a^0_a=2\eta x^a,~a^k_a=2x^kx^a+[\eta ^2-r^2]\delta^k_a).\cr &
~~~~~~~~~~~~~~~~~~~~~~~~~~~~~~~~~~~~~~~~(k=0)}
\m (4.8)
$$

For $k=\pm 1$ the 8 vectors can be written in a more
compact form in terms of the column matrix
$$
\beta={\l({\bf \beta}^{\bullet} \atop {\bf \beta}_{\bullet}\r)} =
{\left(\alpha \atop {\di_0 \alpha}\r)}
~~  {\rm with}~~\alpha= \sin \eta~(k=1)~~ {\rm or} ~~
\alpha= \sinh \eta~(k=-1).
\m (4.9)
$$
What in flat \spttt corresponds to dilatation
and time acceleration can be written as a single combination
(${\bf \delta}$); the same is true of what correspond
to the 3 Lorentz boosts and 3 space accelerations
(${\bf \lambda}_a$)
$$
\eqalign
{&{\bf \delta}={\l({\bf \delta}^{\bullet} \atop {\bf \delta}_{\bullet}\r)} =
(\delta^0=\beta\sqrt{1-kr^2},~
\delta^k=\di_0 \beta \sqrt{1-kr^2}x^k),\cr &
{\bf \lambda}_a={\l({\bf \lambda}^{\bullet} \atop {\bf \lambda}_{\bullet}\r)}=
(\lambda^0_a=\di_0 \beta x^a,~
\lambda^k_a=\beta f^{ka}).}
\m (k= \pm 1)~~~~~(4.10)
$$
Notice that for $k=0$ we can take $\alpha=\eta$ and
apply Eq. (4.10) to that case as well. Then $\beta={\left(\eta
\atop 1\r)}$ and ${\bf \delta}={\left({\bf d}
\atop {\bf t}\r)}$ while ${\bf \lambda}_a={\left({\bf l}_a
\atop {\bf s}_a \r)}$. These 4 vectors are not the same as
those given in Eq. (4.8); only two of
them are in that group ${\bf d}$ and $ {\bf l}_a$,
the other two ${\bf t}$ and
$ {\bf s}_a$ belong to the group of Eq. (4.7).

15 \ckkkk are given by any linear combination
(that are linearly independent of course) with constant
coefficients of Eqs. (4.7) and (4.8) for $k=0$, Eqs. (4.7) and
(4.10) for $k=\pm 1$.

Conformal Killing vectors and their space derivatives appear in the
zero component of the conserved vector densities $\hat{\cal I}^0$ [see
Eq. (4.19) below] in two combinations ${\textstyle{1\over 4}}\bar z$ and a
new one that we denote by  ${\bar y}$:
$$
{\textstyle{1\over 4}}\bar z={\textstyle{1\over 4}}\Bar D_\rho\xi^\rho=
{\textstyle{1\over 3}}\nabla_k {\xi^k}+\dot a \xi^0, ~ ~
\bar y \equiv \l(k + \dot a^2 -\di_0 \dot a\r) \xi^0+{\textstyle{1\over 4}}
(\di_0 \bar z -\dot a \bar z)={\textstyle{1\over 3}}
\nabla^2 {\xi^0}+k \xi^0.
\m (4.11)
$$
Most of the $\bar y$'s are zero. The non-zero one's are
$\bar y({\bf t})=k$ and $\bar y({\bf a}_0)=2$.

There are 3-antisymmetric tensors $\nabla^{[k}\xi^{l]}$ which appear in
the ${\cal I}^{0l}$ components of the superpotential, see Eq. (4.18{\it
a}) or Eq. (4.18{\it b}).  The tensors are not zero for the following
three conformal Killing vectors,
$$
\eqalign
{&\nabla^{[k} s^{l]}_a=-2kx^{[k}s^{l]}_a,~~~\nabla^{[k} r^{l]}_a=
\epsilon^{akl}-2kx^{[k}r^{l]}_a,~~~(k= 0,~\pm 1) \cr &
\nabla^{[k} a^{l]}_a=
4\delta^{[k}_ax^{l]}.~~~~~~~~~~~~~~~~~~~~~~~~~~~~~~~~~~~~~~~~~~(k=0)}
\m (4.12)
$$

\nnn (iii) {\it Superpotentials and conserved vectors for small perturbations.}

We denote the perturbed metric components
$\g.mn$ by $\bar \g.mn+h_{\mu\nu}$. Some authors [42] prefer
to use the ``conformal perturbations" and write
$\g.mn=a^2(e_{\mu\nu}+\~h_{\mu\nu})$. Thus,
$$
ds^2=(\bar \g.mn+h_{\mu\nu})dx^\mu dx^\nu =
a^2(e_{\mu\nu}+\~h_{\mu\nu})dx^\mu dx^\nu.
\m (4.13)
$$
We shall mainly use $\~h_{\mu\nu}$; $h_{\mu\nu}$ seems
to be  preferable in 4-covariant perturbation
calculations. In a $1+3$ splitting, the 10 components
of the perturbations are $\~h_{00},~\~h_{0l},~\~h_{kl}$.
We shall not displace the 0-indices up or down. The $kl$ indices
will
be displaced with the $f_{kl}$ metric. Thus
$$
\~h^m_l=f^{mk}\~h_{kl},~~~\~h^{mn}=f^{mk}f^{nl}\~h_{kl},~~~
\~h^m_0=f^{ml}\~h_{0l}.
\m (4.14)
$$
There are simple relations between $h^\mu_\nu$ and
$\~h_{\mu\nu}$: $h^0_0=\~h_{00},~ h^0_l=\~h_{0l}$ and
$h^k_l=-\~h^k_l$.   The tensor density that enters in the
\sss (3.18) is related to $h_{\mu\nu}$ in the linear
approximation as follows
$$
\hat l^{\mu\nu}=\sqrt{-\bar g}(-\bar g^{\mu\rho}\bar g^{\nu\sig}+{\half}\bar
g^{\mu\nu}\bar g^{\rho\sig})h_{\rho\sig}=
\sqrt{f}(-e^{\mu\rho}e^{\nu\sig}+{\half}e^{\mu\nu}e^{\rho\sig})h_{\rho\sig}.
\m (4.15)
$$
Notice that Eq. (3.18) in terms of $\hat l^{\mu\nu}$ is the same for large
 or small perturbations and can be used for calculations at higher
order  of approximations. In this section, we are interested only in small
perturbations.
There are two 3-tensors that deserve special notations because
they
appear as building blocks of the components of $\hat {\cal I}^{0l}$:
$$
q^m_l \equiv \delta^m_l \~l^{00} - \~l^m_l,
~~Q^m_l \equiv
\nabla^m \~l^{0}_{l}- \di_0 {q^m_l} - \delta^m_l
\l[\nabla^n \~l^{0n} + \dot a (\~l^n_n + \~l^{00})\r],~~\~l^{\mu\nu}
\equiv ({a^2  \sqrt{-\bar g}}) \hat l^{\mu\nu}.
\m(4.16a)
$$
In the linear approximation where Eq. (4.15)
holds the quantities in Eq. (4.16{\it a})
reduce to
$$
q^m_l = \~h^m_l-\delta^m_l\~h^n_n,~~~
  Q^m_l = (2\dot a \~h_{00}-\nabla^n \~h_{0n})\delta^m_l+\nabla^m
\~h_{0l}- \di_0 {q^m_l}.
\m (4.16b)
$$
We also define by a special symbol $\cal Q$  the perturbed
trace of the external curvature of the hypersurface
$\eta=const$ which appear in the zero component of
the conserved vectors. Thus, if $n^\mu$ is the unit normal vector to
that hypersurface,
$$
{\cal Q}\equiv-D_\mu n^\mu-(-\Bar{D_\mu n^\mu}) =
{\textstyle{3\over2}}\dot a \~h_{00}+{\half}\di_0 {\~h}^{n}_n-
\nabla_n \~h_0^n.
\m (4.17)
$$
${\cal Q}=0$ is the ``uniform Hubble expansion" gauge condition which
was introduced by Bardeen [26].

We have now all the elements needed to calculate the conserved vectors
and superpotentials with small perturbations for the 15 \ckkkk and to
write them down in a compact form. We are  particularly  interested in
integral constraints over volumes at a constant time with spherical
boundaries. For this we need only the zero components of the conserved
vector:  $\hat{\cal I}^0=\di_l\hat{\cal I}^{0l}$. Let us  write first
$\hat{\cal I}^{0l}$ which is define by Eq. (3.18).  We obtain after painful
but straightforward calculations the following expression for the
superpotential components which are {\it valid for large perturbations},
using  Eq. (4.16{\it a})
$$
\hat{\cal I}^{0l}={\sqrt{-\bar g}\over 2\k a^2}
\l[(2\dot a\~l^{0l}-\nabla^k q^l_k)\xi^0+
q^l_k \nabla^k {\xi^0} +
Q^l_k\xi^k+
\~l^0_{k}\nabla^{[k}\xi^{l]}\r].
\m (4.18a)
$$
For small perturbations $\~h_{\mu\nu}$
Eq. (4.18{\it a}) reduces with the help of Eq. (4.15) to
$$
\hat{\cal I}^{0l}={\sqrt{-\bar g}\over 2\k a^2}
\l[(2\dot a\~h^l_0-\nabla^k q^l_k)\xi^0+ q^l_k\nabla^k {\xi^0}+Q^l_k\xi^k+
\~h_{0k}\nabla^{[k}\xi^{l]}\r].
\m (4.18b)
$$
The linearized expression for $\hat{\cal I}^0$
is much simpler than it appears in Eq. (3.33) because $\tau^{\mu\nu}
= 0$.
In terms of the energy-momentum perturbations $\delta
T^\mu_\nu=T^\mu_\nu-\bar
T^{\mu}_\nu$ rather than perturbations of densities
 we find that
$$
\hat{\cal I}^0={\sqrt{-\bar g}\over \k a^2}
\l[\k a^2\delta T^0_0 \xi^0+\k a^2\delta T^0_k \xi^k-\bar y
\~h^n_n+{\half}\bar
z{\cal Q}
+\nabla_n(\textstyle{1\over 4}\bar z\~h^n_0)\r].
\m(4.19)
$$
Equation (4.19) suggests that it makes sense to transfer
$\nabla_n(\textstyle{1\over 4}\bar
z\~h^n_0)$ from the left to the right hand side in
$\hat{\cal I}^0=\di_l\hat{\cal I}^{0l}$
and to rewrite it in the
following  ``renormalized'' form appropriate to a 3-dimensional formalism:
$$
*{\cal I}=\nabla _l\l(*{\cal I}^l\r), ~~~~~~
*{\cal I}\equiv {\k a^2\over \sqrt{-\bar g}}\hat{\cal
I}^0- \nabla_l({\textstyle{1\over 4}}\bar z\~h^l_0),
~~~*{\cal I}^l \equiv {{\k a^2}\over {\sqrt{-\bar g}}}\hat{\cal
I}^{0l}-{\textstyle{1\over 4}}\bar z\~h^l_0.
\m (4.20)
$$
The zero indices are no more appropriate
because the stared quantities are not
the components of conserved vectors or superpotentials anymore.
With the definitions in Eq. (4.20) we can write instead of Eq. (4.19)
$$
*{\cal I}= \Pi_0 \xi^0+\Pi_k \xi^k -\bar y
\~h^n_n+{\half}\bar z{\cal Q},~~~~~~~
\Pi_\mu\equiv\k a^2\delta T^0_\mu,
\m (4.21)
$$
and instead of (4.18{\it b})
$$
*{\cal I}^l={1\over 2}\l(-\nabla^k q^l_k\xi^0 + q^l_k\nabla^k{\xi^0}
+Q^l_k\xi^k -{\textstyle{2\over 3}}\~h^l_0\nabla_k{\xi^k}
+\~h_{0k}\nabla^{[k}\xi^{l]}\r).
\m (4.22)
$$
Integrating
$*{\cal I}=\nabla_l\l(*{\cal I}^l\r)$
over a sphere $ (r=const)$ at constant time $\eta=const$ we obtain
$$
\int_\eta  *{\cal I} \sqrt{f}d^3x = r^2\sqrt{f}\oint_r
\delta_{kl}\l(*{\cal I}^k\r) {x^l\over
r}sin(\theta)d\theta d\phi.
\m (4.23)
$$

We now give the list of the  15 ${*\cal I}$'s and their associated
${*\cal I}^l$'s.
Some linear combinations with $\eta$ dependent factors have greater
simplicity. Such combinations break of course the group character of
the algebra of globally conserved quantities but here we are interested
in integral constraints at a given time $\eta$ for which  the group
properties of  our currents are not important here. We shall keep
trace however of
the corresponding $\eta$-dependent combinations of \ckkkk
and use special symbols for $*{\cal I}$ and $*{\cal I}^l$  that reminds
us of their origin.  For instance $*{\cal I}({\bf t})$ is denoted by
${\cal T}$, $*{\cal I}({s_a})$ by ${\cal S}_a$ and so on... . Thus
for $k=0$ and $k=\pm 1$ we have the following $*{\cal I}$'s and $*{\cal
I}^l$'s
with $*{\cal I}=\nabla_l \l(*{\cal I}^l\r)$:
$$
{\bf t}  \rightarrow {\cal T}=\nabla_l{\cal T}^l, ~~~~~~~~
{\cal T}=\Pi_0+2\dot a{\cal Q}-k\~h^n_n={\cal T}_0-k\~h^n_n,
~~~~{\cal T}^l=-{\half}\nabla^k q_k^l,~~~~~~~~~~~~~
\m (4.24~I)
$$
$$
{\bf s}_a  \rightarrow {\cal S}_a=\nabla_l{\cal S}_a^l,~~~~
{\cal S}_a= \Pi_a \sqrt{1-kr^2}, ~~~~
{\cal S}_a^l= {\half}Q^l_a\sqrt{1-kr^2}-k\~h_{0k}x^{[k}s^{l]}_a,~~~~
\m (4.24~II)
$$
$$
{\bf r}_a  \rightarrow {\cal R}_a=\nabla_l{\cal R}_a^l,~~{\cal R}_a=
\Pi_k \epsilon_{kan}x^n,~~
 {\cal R}_a^l={\half}(Q^l_k\epsilon_{kan}x^n+\~h_{0k}\epsilon^{akl})-
k\~h_{0k}x^{[k}r^{l]}_a.~~~~~~~~~
\m (4.24~III)
$$
The following 8 quantities in which appears ${\cal T}_0$ defined in Eq.
(4.24$~$I) are for $k=0$ only:
$$
{\bf l}_a-\eta{\bf s}_a \rightarrow {\cal L}_a=\nabla_l{\cal L}_a^l,~~~~
{\cal L}_a= {\cal T}_0x^a,~~ ~~
{\cal L}_a^l=-{\half}\nabla^k q_k^l x^a + \half q^l_a,~~~~~~~~~~~~~~~
\m (4.24~IV)
$$
$$
{\bf d} - \eta {\bf t}\rightarrow {\cal D}=\nabla_l{\cal D}^l,~~~~
{\cal D}={\cal S}_ax^a+2{\cal Q},~~~~
{\cal D}^l= {\half}Q^l_k x^k-\~h^{0l},~~~~~~~~~~~~~~~~~~~~~~
\m (4.24~V)
$$
$$
{\bf a}_o + \eta^2 {\bf t}-2\eta{\bf d}\rightarrow
{\cal A}_0=\nabla_l {\cal A}_0, ~~~~{\cal A}_0={\cal T}_0r^2-2\~h^n_n,~~~~
{\cal A}_0^l=-{\half}\nabla_k q^{kl} r^2+q^l_kx^k,
\m (4.24~VI)
$$
$$
\eqalign
{&{\bf a}_a+\eta^2{\bf s}_a-2\eta{\bf l}_a\rightarrow \cr &
{\cal A}_a=\nabla_l{\cal A}_a^l,~{\cal A}_a=2{\cal D}x^a- {\cal S}_ar^2,
{\cal A}_a^l=Q^l_k(x^kx^a-{\half}r^2\delta^k_a) +
2\~h_{0k}\l(\delta^{[k}_ax^{l]} - \delta^{kl} x^a\r).~~~~~~~}
\m (4.24~VII)
$$
The next 8 linear combinations of \ckkkk are for $k=\pm 1$ only. In
those
formulas the expressions like ${\bf\lambda}_a(\beta)$
and ${\bf\lambda}_a(\di_0\beta)$ represent the
factors of $\beta$ and $\di_0\beta$ in the conservation law associated
with
${\bf\lambda}_a$. Thus
$$
{\bf\lambda}_a(\di_0\beta) \rightarrow \~{\cal L}_a=\nabla_l \~{\cal L}_a^l,~~
\~{\cal L}_a={\cal T}_0x^a,~~~~
\~{\cal L}_a^l=-{\half}\nabla^k q_k^l x^a+{\half}q^{l}_{a},
~~~~~~~~~~~~~~~~~~~~~~~
\m (4.24~\tilde{IV})
$$
$$
{\bf \delta}(\di_0 \beta) \rightarrow \~{\cal D}=\nabla_l\~{\cal D}^l,~~
\~{\cal D}={\cal S}_ax^a+2{\cal Q}\sqrt{1-kr^2},~~
\~{\cal D}^l= ({\half}Q^l_k x^k -\~h^l_0)\sqrt{1-kr^2},~~~
\m (4.24~\tilde V)
$$
$$
 {\bf \delta}(\beta) \rightarrow \~{\cal A}_0=\nabla_l\~{\cal A}_0^l,~~
\~{\cal A}_0={\cal T}_0\sqrt{1-kr^2},~
\~{\cal A}_0^l=-{\half}(\nabla_k q^{kl} +kq^l_kx^k)\sqrt{1-kr^2},~~
\m (4.24~\tilde {VI})
$$
$$
{\bf\lambda}_a(\beta) \rightarrow \~{\cal A}_a=\nabla_l\~{\cal A}^l_a,~~
\~{\cal A}_a=\Pi_kf^{ka}-2kx^a{\cal Q},~~
 \~{\cal A}_a^l={\half}Q^l_a\eta+k\~h^l_0x^a.~~~~~~~~~~~~
\m (4.24~\tilde {VII})
$$
We notice that $~\~{\cal D}(k=0)={\cal D}~$ and
$~\~{\cal D}^l(k=0)={\cal D}^l~$. Also
$~\~{\cal L}_a(k=0)={\cal L}_a$ and
$~\~{\cal L}^l_a(k=0)={\cal L}^l_a$. However,
$\~{\cal A}_0(k=0)\neq{\cal A}_0$ and
$\~{\cal A}^l_0(k=0)\neq{\cal A}^l_0$ as well as
 $\~{\cal A}_a(k=0)\neq {\cal A}_a$.
and $\~{\cal A}^l_a(k=0)\neq {\cal A}^l_a$.

\nnn (ii) {\it Analysis of these results.}

${\cal T}$, ${\cal S}_a$, ${\cal R}_a$ and ${\cal L}_a$, ${\cal D}$, ${\cal A}_0$, ${\cal A}_a$ for $k=0$ or
$\~{\cal L}_a$, $\~{\cal D}$, $\~{\cal A}_0$, $\~{\cal A}_a$ for $k=\pm 1$ contain three types of terms: linear
combinations of $a^2\k \delta T^0_\mu=\Pi_\mu$, the perturbation of the``uniform Hubble expansion" ${\cal Q}$ and the
trace of the perturbation of the spatial components of the metric $\~h^n_n$.

${\cal S}_a$ and ${\cal R}_a$ are homogeneous in $\Pi_\mu$.
 Besides these 6 quantities there are 4 additional linear combinations
with
$\eta$-dependent coefficients that are homogeneous in $\Pi_\mu$. For $k=0$
$$
\eqalign
{&\dot a^{-1}{\cal L}_a-{\half}{\cal A}_a  =
\dot a^{-1}\Pi_0x^a-\Pi_k(x^kx^a-{\half}\delta ^{ka}r^2)\equiv \Pi_\mu
V^\mu_a, \cr &
\dot a^{-1}{\cal T}-{\cal D}  =
\dot a^{-1}\Pi_0-\Pi_kx^k\equiv \Pi_\mu V^\mu_0.}
\m (k=0)~~~~~(4.25)
$$
For $k=\pm 1$
$$
\eqalign
{&\dot a^{-1}\~{\cal L}_a+k\~{\cal A}_a  =
\dot a^{-1}\Pi_0x^a+k\Pi_kf^{ka}\equiv\Pi_\mu\~V^\mu_a, \cr &
\dot a^{-1}\~{\cal A}_0-\~{\cal D}  =
(\dot a^{-1}\Pi_0-\Pi_kx^k)\sqrt{1-kr^2}\equiv \Pi_\mu \~V^\mu_0.}
\m (k = \pm 1)~~~~~(4.26)
$$
The $V$'s and $\~V$'s are Traschen's [1] ``integral constraint vectors".
Her vectors are thus linear combinations of conformal Killing vectors
with time dependent coefficients.
In particular the $\~{\bf V}_a$'s are linear combinations of
${\bf\lambda}_a(\di_0 \beta)$'s and ${\bf\lambda}_a(\beta )$'s while
$\~{\bf V}_0$ is a combination of ${\bf\delta}(\di_0\beta )$'s and
${\bf\delta}(\beta )$.

It is however clear that if we take the uniform
Hubble expansion gauge condition
$$
{\cal Q}=0,
\m (4.27)
$$
then  14 of the 15 ``conservation laws" have volume
integrands that are linear and homogeneous in $\Pi_\mu$. The exception is
for the conformal time translation ($k=\pm1$) or acceleration ($k=0$):
$$
 {\cal T}=\Pi_0-k\~h^n_n,  ~~~(k=\pm 1);~~~~~~~~
{\cal A}_0=\Pi_0r^2-2\~h^n_n,~~~(k=0).
\m (4.28)
$$

Thus if ${\cal Q}=0$, the conformal Killing vectors provide 14 linearly
independent expressions that are momenta of order 0, 1 and 2 and are
given by surface integrals involving boundary values only. Such
expressions can in principle be constructed directly from Einstein's
constraint equations.  The constructs are however far from obvious.

\nnn (iii) {\it Example: \sptttt that are asymptotically
Schwarzschild-de Sitter $(k=0)$.}

In  this example, the background is a de Sitter \spttt with $k=0$ and
perturbations far away from its sources appear to be spherically
symmetrical. However, perturbations may be
large at and near the sources. The asymptotic metric in our coordinates
has been given by Bi\v c\'ak and Podolski [43] [their formula (48)].
Neglecting powers of $m/r$ higher than one, their metric is as follows
$$
ds^2=d\tau^2-(e^{2H\tau}+2F)d\chi^2 - (e^{2H\tau}-F)\chi^2
(d\theta^2+sin^2(\theta)d\phi^2),~~~~
F\equiv{2m\over\Lambda}\chi^{-3}e^{-H\tau}
\m (4.29)
$$
and $H=({\Lambda / 3})^{1/2}$; $m$ and $\Lambda$ are constants. We set
$$
\eta=\varepsilon e^{-H\tau}, ~~~r= H\chi, ~~~~\varepsilon \equiv \pm 1
\m (4.30)
$$
so that Eq. (4.29) takes no this form
$$
ds^2=(H\eta)^{-2}[d\eta^2 +(-\delta_{kl}+\~h_{kl})dx^kdx^l].
\m (4.31)
$$
Thus, comparing to Eq. (4.13), we see that
$$
a=(H|\eta|)^{-1},~~ \~h_{00}=0,  ~~\~h_{0l}=0,~~
\~h_{kl}=-\varepsilon {\textstyle{2 \over 3}}mH
\l({\eta\over r}\r)^3\l(\delta_{kl}- {{3\delta_{km}x^m\delta_{ln}x^n}
\over {r^2}}\r).
\m (4.32)
$$
Notice that ${\cal Q}=0$ in this example and furthermore $\~h^n_n=0$.
To calculate $*{\cal I}^l$ we need the $q$'s and $Q$'s defined in Eq.
(4.16{\it b}): $q^l_k=\~h^l_k$ and $Q^l_k=-({3 / \eta})\~h^l_k$. With these
elements we find that all 13 integrals that follow are zero:
$$
\int_\infty  {\cal T} \sqrt{f}d^3x=\int_{\infty}  {\cal S}_a
\sqrt{f}d^3x=\int_{\infty}{\cal R}_a \sqrt{f}d^3x=\int_{\infty}{\cal L}_a
\sqrt{f}d^3x=\int_{\infty}{\cal A}_a \sqrt{f}d^3x=0.
\m (4.33)
$$
The equalities constitute as many Traschen-like integral constraints.
The 2 constants of motion that are not equal to zero are associated with
dilatations and with time accelerations in Minkowski spacetime which is
conformal to our background
$$
-{\varepsilon \over \k\eta^2}\int_{\infty}  {\cal D} \sqrt{f}d^3x = HMc^2,
 ~~~~{\varepsilon\over
\k\eta^3}\int_{\infty}  {\cal A}_0 \sqrt{f}d^3x =
{\textstyle{2 \over 3}}HMc^2.
\m (4.34)
$$
In these expressions $M={mc^2 / G}$. Notice that the 15 integrands
of Eqs. (4.33) and (4.34) do not have to be linearized. Near the sources
perturbations may be large. If the perturbations are weak in
the whole space, we may write instead of Eq. (4.34)
$$
-{\varepsilon\over \eta^2}\int_{\infty}
\delta T^0_kx^k \sqrt{f}d^3x = HMc^2, ~~~~{\varepsilon\over
\eta^3}\int_{\infty}  \delta T_0^0r^2 \sqrt{f}d^3x =
{\textstyle{2 \over 3}}HMc^2.
\m (4.35)
$$
These are integral
constraints on $\delta T^0_0$ and $\delta T^0_k$.

\beginsection 5. Comments on the role of superpotentials in the
theory of conservation laws

\noindent (i) {\it Motivations.}

Here we want to connect our work with past literature, give due credit to
yet unmentioned
papers and make contact with some  well known superpotentials or
energy-momentum tensors that
we have not  yet encountered.

\noindent (ii) {\it On superpotentials in conservation laws today.}

Perhaps the single most important legacy of studies on conservation
laws in general relativity is that conserved quantities in finite
volumes can always be expressed as surface integrals on the boundary of
the volume. Anti-symmetric tensor densities like $\hat {\cal
I}^{\mu\nu}$ dominate the scene to day in the literature, not $\hat
{\cal I}^\mu$.  One great push in that direction was given by Penrose
[14] who introduced the notion of ``quasi-local'' quantities which, in
the weak field limit, reduce to ordinary conserved linear and angular
momentum of the gravitational field in finite volumes.  Many papers
have been published on the subject in particular on quasi-local energy.
Unfortunately, selection rules are few and no consensus exists.  There
is an interesting comparison of formulas in a paper by Berqvist  [44]
on the energy enclosed by the outer horizon of a Kerr black hole in
which it is shown how six different formulas give five different
results.  There exists however a common point to those various
definitions of quasi-local energy: they are not related by differential
conservation laws to Einstein's equations [45]. In this instance, the
present work points in a very different direction. We have not tried to
make the connection with our own superpotential. The role and
importance of \ssss in field theory has been emphasized by Julia and
Silva [28] who gave them an elegant and general mathematical basis.

\noindent (iii) {\it Connection with other superpotentials on a flat
background.}

Rosen [19] was the first to drive attention to the fact that the quadratic
$\Gamma\Gamma$-Lagrangian, Eq. (2.6) used by Einstein to derive a conserved
pseudo-tensor could be written in covariant form by introducing a
second metric. This amount in practice to describe curved \sptttt with
respect to a flat background.
The mathematical basis of Rosen's approach
is given in Lichnerowiscz [46].
It is thus no surprise that our
formulation of conservation laws for perturbations of curved
backgrounds   connects nicely with well known conservation laws in
classical general relativity. This is what we want to show here.  Let
us go back for a moment to the ``divergence dependent'' conserved vector
$\hat I^\mu$ with Eq. (2.17), not $\hat {\cal I}^\mu$. The Rosenfeld
identities have been worked out in KBL97\footnote {****}{In KBL97, the
formulas are (2.50) to (2.52) but a $~\hat{}~$ is missing on every
symbol!}.  On a flat background, $\Bar R^\lambda ~_{\nu\rho\sig}=0$ and
in arbitrary coordinates, the formulas of KBL97 are similar to Eqs.
(3.27) - (3.30).  In particular,
$$
 \Bar D_\mu \hat \theta^\mu_\nu  =0,~~~~ \hat \theta^{\mu\nu} = -\Bar
D_\lambda \hat
\sig^{\lambda\mu\nu},~~~~
\hat \theta^\mu_\nu=\hat T^\mu_\nu+\hat t^\mu_\nu,
\m(5.1)
$$
$\hat t^\mu_\nu$ is
Einstein's energy-momentum {\it tensor} density
as given by Rosen  in
arbitrary coordinates.
On a flat background $\hat \theta^\mu_\nu$ is thus the divergence of a
tensor, not an
anti-symmetric one and not a two index tensor but still one acting like a
``superpotential'' for
volume integrals in Minkowski coordinates.
$\hat
\sig^{\lambda\mu\nu}$ is Tolman's [47] superpotential, apparantly the
first
of its kind in the
literature. This superpotential is related to another famous
one,  Freud's [33]
superpotential $\hat F^{\lambda\mu\nu}$ is defined in arbitrary
coordinates by Eq. (2.22):
$$
\hat \sig^{\lambda\mu\nu}=\hat F^{\lambda\mu\nu}+ {1\over
2\k}
\Bar
D_\rho\l(\bar g^{\nu[\rho} \hat l ^{\lam]\mu}\r).
\m(5.2)
$$
Since covariant derivatives on a flat spacetime are commutative,
by taking the divergence of $\hat \sig ^{\lambda\mu\nu}$ and using its
relation with $\hat \theta ^{\mu\nu}$ - see Eq. (5.1) - we obtain a
similar relation
$$
\hat \theta^{\mu\nu}=-\Bar D_\lambda \hat
F^{\lambda\mu\nu}.
\m (5.3)
$$
The great simplicity of Freud's superpotential made it a successful
quantity to calculate globally conserved quantities like the total
energy at spatial infinity as well as at null infinity [48].  Our own
energy tensor on a flat background satisfies similar relations.
$$
\Bar D_\mu \hat {\cal T}^\mu_\nu =0,~~~~
 \hat {\cal T}^{\mu\nu} = -\Bar D_\lambda(*{\hat S}^{\lambda\mu\nu})=
-\Bar D_\lambda \hat {\cal P}^{\lambda\mu\nu}.
\m (5.4)
$$
The difference between  $\hat {\cal T}^{\mu\nu}$ and $\hat
\theta^{\mu\nu}$
following from Eq. (3.10) is exactly $\Bar D_\rho \hat S^{\rho\mu\nu}$. It is
interesting
to notice the relation between the divergences of the $*S$ and $\cal
P$ tensors
when  backgrounds are not flat:
$$
\Bar D_\lambda \hat {\cal P}^{\lambda\nu\mu} =
\Bar D_\lambda (*\hat S^{\lambda\mu\nu}) -
{1\over 2\k} \hat l^{\rho\sig}{\Bar R^{\mu}_{~\rho\sig}}^{\nu} -
{1\over 2\k} \hat l^{\lambda\nu}\Bar R^{\mu}_\lambda.
\m(5.5)
$$

A most famous superpotential is that of Komar,
$(1/\k)D^{[\mu}\hat \xi^{\nu]}$.  Its
greatest quality is to be background independent. It is also useful in asymptotically flat
spacetimes at spatial infinity.
But  it has some  shortcomings which we already mentioned. There
have been
various corrections of
that attractive covariant expression [49]
which did not get rid of the anomalous factor 2
and had also other ``defects"
[45].

One intriguing superpotential is that of Arnowitt, Deser and Misner
[50], especially for energy. It was original defined in a synchronous
gauge, at least asymptotically, $g_{00}=1$, $g_{0k}=0$.  In  that gauge,
the surface integral at spatial infinity of the Komar tensor is zero.
What remains then of the superpotential in Eq. (2.21) is  the
(0k)-component of $(1/\k)\xi^{[\mu}\hat k^{\nu]}$ which reduce indeed
to the ADM integrand at infinity as can easily be verified.

The Landau and Lifshitz  superpotential is as satisfactory as our
own superpotential for calculating the total 4-linear momentum but it
has the wrong weight and it is difficult to see how to connect it with
the group of diffeomorphisms via N\oe ther's method on a curved
background.  The L-L complex has however been obtained recently from a
variational principle by Babak and Grishchuk [51] on a flat
background in arbitrary coordinates.  Incidentally L-L's pseudo-tensor
has one more drawback, not shared by Einstein's pseudo-tensor
which was pointed out by Chandrasekhar and Ferrari [52]. Consider the
weak field approximation of the total energy in a stationary
spacetime. A variational principle applied to the total ``Einstein
Energy'' leads to Einstein's linearized field equations. The ``L-L
Energy'' provides incorrect equations.

\vskip .2 in

\beginsection {\bf Acknowledgments}

We are very grateful to Nathalie Deruelle's critical review of the
summary presented at the Conference on ``Fundamental Interactions:
from Symmetries to Black-holes"
IN Brussels (March 1999) in honor of
Fran\c cois Englert. Her remarks helped to greatly improve
the present expanded version. We thank Chiang-Mei Chen for a useful
explanation of his work with James Nester and we had a very interesting
correspondence with James Nester himself. Alexander Petrov had
profitable exchanges with Piotr Chru\'sciel, Ji\v r${\acute \imath}$
Bi\v c\'ak and Leonid
Grichshuk which he acknowledges with thanks. Many thanks also to Joshua
Goldberg who read the summary and made precious comments. Joseph Katz
enjoyed, as usual, the illuminating comments and deep views of Donald
Lynden-Bell and of Ji\v r$\acute \imath$ Bi\v c\'ak to whom he is ever grateful.

\beginsection Appendix on
Globally Conserved Quantities

\noindent {\it (i) Object of this appendix.}

There exists quite a number of different conservation laws but few
criteria to  select among them.
The main discriminating conditions are global conservation laws and the
weak field limit which
are to some extend related. The only non-ambiguously defined global
quantities are the 4
components of the linear momentum
$P_\alpha$ at spatial and at null infinity for spacetimes with definite
fall off conditions to
asymptotic flatness.  It is therefore important to show that our
 superpotential gives at least the correct total 4-momentum in
those cases. In this task we shall be
greatly helped by Eq. (3.11) which for \kkk is
$$
\hat
 {\cal J}^{\mu\nu}= \hat
  J^{\mu\nu}+\hat {S}^{\mu\nu\rho}\bar\xi_\rho.
\m (A.1)
$$
It has been shown  that the \sss $\hat
  J^{\mu\nu}$ provides by itself the total 4-momentum $P_\alpha$ both at
spatial infinity [16] and at null infinity [53]. We must therefore show
that $\hat
{S}^{\mu\nu\rho}\bar\xi_\rho$ does not contribute to $P_\alpha$ in both
asymptotic directions.
This is what we briefly indicate in this appendix. We shall show that in both
asymptotic
directions the
``discrepancy"
$$
\Delta P_\alpha = \oint_S \hat {S}^{\mu\nu\rho}\bar\xi_{(\alpha)\rho}
dS_{\mu\nu}=0.
\m(A.2)
$$
$S$ is the sphere at infinity and $\bar\xi_{(\alpha)\rho}$,
$ (\alpha = 0,1,2,3)$
are the four Killing vectors of translations in the asymptotically
flat background.

\noindent {\it (ii) Stationary \sptttt and spatial infinity.}

Consider stationary solutions that fall off as follows in asymptotic
Minkowski coordinates at spatial infinity $x^\mu = (x^0=t, x^k)$:
$$
g_{\mu\nu}(x^k)=\eta_{\mu\nu}+ {1\over r}u_{\mu\nu}+O_2,~~~~ ~~~~\di_l
g_{\mu\nu}(x^k)={1\over r^2}v_{l\mu\nu}+O_3
\m (A.3)
$$
\noindent where $r=\sqrt{\Sigma(x^k)^2}$ and $u_{\mu\nu},~ v_{l\mu\nu}$ depend  on the
directions on the sphere
at infinity. The background metric is
$\eta_{\mu\nu}$  and the
\kkk of translations $\bar \xi_{(\alpha)\nu}=\eta_{\alpha\nu}$. The
discrepancy is therefore
given by
$$
\Delta P_\alpha = \oint_{r\rightarrow\infty}  \hat {S}^{0l\alpha}n_l~r^2
sin\theta d\theta
d\phi, ~~~~~~n_l={X^l\over r}.
\m(A.4)
$$
It takes some patience to make this calculation, using the asymptotic
form of the metric (A.3) in Eq. (3.8) with Eq. (3.21),
but the calculation does
not need clever tricks and we find  indeed that $\Delta P_\alpha =0$
for $r\rightarrow\infty$. The metric does not have to fall off as fast
as in Eq. (A.3).  $\Delta P_\alpha =0$ under weaker fall off
conditions has been studied in [54].

\noindent {\it (iii) Radiation \ssss at null infinity.}

For the Bondi-Sachs metric we use the Newman and Unti [55]
representation with coordinates
$x^\lambda =(x^0=u,~  x^1=r,~ x^2,~x^3)
$ in which the metric has the following form
$$
ds^2= g_{00}du^2+2dudr+2g_{0L}dudx^L+g_{KL}dx^Kdx^L
\m (A.5)
$$
and the flat background has a
$$
d\bar s^2= du^2+2dudr-{r^2\over 2P^2}[(dx^2)^2+(dx^3)^2],~~~~~~
P=\half+\txt{1\over 4}[(x^2)^2+(x^3)^2].
\m(A.6)
$$
The asymptotic form of the metric depends on 5 independent real
functions of $(u,x^2,x^3)$, namely $\psi'_1$, $\psi''_1$, $\psi'_2$,
$ \sig'$ and $\sig''$. These notations are similar to those of Newman
and Unti - without a   zero index - who use complex  functions; here
the  prime indicates the real part, two primes indicate the complex
part of their complex functions. The asymptotic form of the metric in
these notation is
$$
g_{00}=1 + {2\psi'_2 \over r} + O_2,
\m(A.7)
$$
$$
g_{02}=-P^2[\di_2({\sig '\over P^2})+\di_3({\sig ''\over P^2})] +
{2\over 3}{{\psi'}_1 \over {Pr}} + O_2,~~~
g_{03}=
{P^2}[\di_3({\sig ''\over P^2})-\di_2({\sig '\over P^2})] +
{2\over 3}{{\psi''_1} \over {Pr}} + O_2,
\m(A.8)
$$
$$
g_{23}     = -r{\sig''\over P^2}+O_1,~~~
g_{22}     = - {r^2 \over 2P^2} - {r\sig' \over P^2}
         +O_0,~~~
g_{33}   = - {r^2 \over 2P^2} + {r\sig' \over P^2} + O_0.
\m(A.9)
$$
with $|\sig|^2= \sig '^2+\sig''^2$.

The \kkk of translations
in the flat background spacetime with the metric (A.6)
have the following components
$$
\bar \xi_{(0)\mu}  = (1,~1,~0,~0),~~~
\bar \xi_{(m)\mu}=\l(0,~-n_m,~-r\di_2 n_m,~-r\di_3 n_m\r),~~~
n_m = {X^m \over r}.
\m(A.10)
$$
With Eq. (A.5) to Eq. (A.10), we have the necessary elements to calculate
Eq. (A.2) at null infinity using Eq. (3.8) with Eq. (3.21).
The calculation is  even
more tedious than before
but still it is not difficult to show that indeed
$\Delta P_\alpha =0$ for $r\rightarrow\infty$.

The reader may remember that the loss of energy $E$ per unit u-time is
given
by
$$
{dE\over du} = -(8\pi)^{-1}\oint [(\di_u \sig ')^2 +(\di_u \sig
'')^2]P^{-2}
dx^2dx^3 < 0.
\m (A.11)
$$
This formula is one of the outstanding results of Bondi.

The loss of energy obtained from the Abbot and Deser [37]
superpotential is different and is not negative definite.  Following
the prescription indicated in section 3 this \sss satisfy the following
expression
$$
\k J^{\mu\nu}_{AD} =
- H^{\rho[\mu}\Bar D_\rho\bxi^{\nu]} +
\bxi^\rho\Bar D^{[\mu} H^{\nu]}_\rho -
\bxi^{[\mu}\Bar D_\rho H^{\nu]\rho}
\m(A.12)
$$
where
 $$
H_{\mu\nu} = *h_{\mu\nu} - \half \Bar g_{\mu\nu}(*h),~~~~
  *h_{\mu\nu}=g_{\mu\nu}-\Bar g_{\mu\nu},~~~~*h=\bar
g^{\rho\sig}(*h_{\rho\sig}).
\m(A.13)
$$
If we use this superpotential to calculate the energy $E_{AD}$ and the
corresponding energy
loss, we find that
$$
{dE_{AD}\over du} = {dE\over du} + (8\pi)^{-1}{d^2\over du^2}
\oint {|\sig|^2\over P^2} dx^2dx^3.
\m (A.14)
$$
This is not negative definite. The Abbott and Deser superpotential
should be rejected on this basis.

\vskip .3 in

\centerline {BIBLIOGRAPHY}
\vskip .2 in
\nnn [1] J. Traschen, Phys. Rev. D {\bf 31}, 283 (1985).

\nnn [2] J. Traschen  and D.M. Eardley, Phys. Rev. D {\bf 34}, 1665 (1986).

\nnn [3] R.K. Sachs and A.M. Wolfe, Ap. J. {\bf 147}, 73 (1967).

\nnn [4] G.F.R. Ellis  and M. Jaklitsch,  Ap. J. {\bf 346}, 601 (1989);
W.R. Stoeger, G.F.R. Ellis and B.G. Schmidt, Gen. Relat. Grav.
{\bf 23}, 1169 (1991).

\nnn [5] K.P.Tod, Gen. Relat. Grav. {\bf 20}, 1297 (1988).

\nnn [6] J. Katz, J. Bi\v c\'ak and D. Lynden-Bell,  Phys.
Rev. D {\bf 55}, 5759 (1997).

\nnn [7] S. Veeraraghavan and A. Stebbin, Ap. J. {\bf 365},  37 (1990).

\nnn [8] J.P. Uzan, N. Deruelle and N. Turok,  Phys. Rev. D {\bf 57},
7192 (1998).

\nnn [9] J. Katz, J. Bi\v c\'ak and D. Lynden-Bell,
 Monthly Notices Roy. Astr. Soc. {\bf 272}, 150 (1995); Erratum
{\bf 277}, 1600 (1995).

\nnn [10] L.D. Landau  and E.M. Lifshitz, {\it The Classical Theory of
Fields} (Pergamon Press, London, 1951).

\noindent [11] P.G. Bergmann  and R. Schiller, Phys. Rev. {\bf 89}, 4 (1953).

\noindent [12] A. Trautman, in {\it Gravitation: an Introduction to
Current Research},
Ed. L. Witten (Wiley, New York, 1962).

\noindent [13] A. Komar, Phys. Rev. {\bf 113}, 934 (1959).

\noindent [14] R. Penrose,  Proc. R. Soc. (London) A {\bf 381}, 53 (1982).

\noindent [15] H. Bondi, Nature {\bf 186}, 535 (1960).

\noindent [16] J. Katz,  Class. Quantum Grav. {\bf 2}, 423 (1985).

\noindent [17] C. M\o ller,  Ann. Phys. {\bf 12}, 118 (1961).

\noindent [18] J.W. York,{\it Jr.}, Gen. Rel. Grav. {\bf 18}, 249 (1986).

\noindent [19] N. Rosen, Phys. Rev. {\bf 57}, 147 (1940).

\noindent [20] F. Belinfante, Physica {\bf 6}, 887 (1939).

\noindent [21] A. Papapetrou  Proc. R. Irish Ac. {\bf 52} 11 (1948)

\noindent [22] D. Bak, D. Cangemi and R. Jackiw, Phys. Rev. D {\bf 49},
5173 (1994).

\noindent [23] C.C. Chang, J.M. Nester and C.M. Chen,  Phys. Rev. Lett. {\bf
83} 1897 (1999); C.M. Chen and J.M. Nester, Class. Quantum Grav.
{\bf 16}, 1279 (1999).

\noindent [24] A.Z. Petrov,  {\it Einstein spaces}
(Pergamon Press, London, 1969).

\noindent [25] T. Fulton, F. Rohrlich and  L. Witten, Rev. Mod. Phys.
{\bf 34}, 442 (1976).

\nnn [26] J.M.Bardeen, Phys. Rev. D {\bf 22}, 1882 (1980).

\noindent [27] L. Rosenfeld, Acad. R. Belg. Mem. Class. Sci.
{\bf 18}, 1 (1940).

\nnn [28] B. Julia and S. Silva,  Class. Quantum Grav. {\bf 15}, 2173 (1998).

\noindent [29] A. Einstein, Sitzungsber. preuss. Akad. Wiss.
{\bf 2}, 1111 (1916).

\noindent [30] S.N. Gupta,  Proc. R. Soc. (London) A
{\bf 65}, 161 (1952).

\noindent [31] N.N. Bogoliubov and D.V. Shirkov,  {\it Introduction to the
theory of quantized fields} (Interscience, New York, 1959).

\noindent [32] P.G. Bergmann,  Phys. Rev. {\bf 75}, 680 (1949).

\noindent [33] P. Freud, Ann. Math. (Princeton) {\bf 40}, 417 (1939).

\noindent [34] F.H.J. Cornish, Proc. R. Soc. (London) A
{\bf 282}, 358; 372 (1964).

\noindent [35] S. Weinberg,  {\it Gravitation and Cosmology}   (John Wiley,
New York, 1972).

\noindent [36] C.W. Misner, K.S. Thorne  and  J.A. Wheeler,  {\it Gravitation}
(Freeman, San Fransisco, 1973).

\nnn [37] L.F. Abbott  and S. Deser, Nuclear Phys. B {\bf 195}, 76 (1982).

\noindent [38] R.K. Sachs,  Proc. R. Soc. (London) A {\bf 270}, 103 (1962).

\noindent [39] H. Bondi, M.G.J. van der Burg  and A.W.K.  Metzner,
Proc. R. Soc. (London) A {\bf 269}, 21 (1962).

\noindent [40] L.P. Grishchuck , A.N. Petrov  and A.D. Popova,
Commun. Math. Phys. {\bf 94},  379 (1984).

\noindent [41] R. Penrose and W. Rindler, {\it Spinors and Space-Time:
Spinor and Twistor Methods in Space-Time Geometry} (Cambridge
Univ. Press, Cambridge, 1988).

\nnn [42] E. Bertschinger, in {\it Cosmology and large scale structures},
Eds. R. Schaefer, J. Silk, M. Spiro and J. Zin-Justin
(North Holland, Amsterdam, 1996), p. 273.

\nnn [43] J. Bi\v c\'ak and J. Podolsk\'y, Phys. Rev. {\bf 52}, 887 (1995).

\noindent [44] G. Bergqvist, Class. Quantum Grav. {\bf 9}, 1753 (1992).

\noindent [45]  J.N. Goldberg, Phys. Rev. D  {\bf 41}, 410 (1990).

\noindent [46] A. Lichnerowicz,  {\it Th\'eories relativistes de la
Gravitation et de l'Electromagn\'etisme} (Masson, Paris, 1955).

\noindent [47] R.C. Tolman,  {\it Relativity, Thermodynamics and
Cosmology} (Oxford
Univ. Press, London, 1934).

\noindent [48]  J.N. Goldberg, Phys. Rev. D {\bf 131}, 1367 (1963).

\noindent [49] L. Tamburino and J. Winicour,  Phys. Rev.
{\bf 150}, 1039 (1966);  R. Geroch and J. Winicour,  J. Math. Phys.
{\bf  22}, 803 (1981).

\nnn [50]  R. Arnowitt, S. Deser and C.W. Misner,
Phys. Rev. {\bf 118}, 1100 (1960).

\noindent [51] S.V. Babak  and  L.P. Grichshuk, {\bf gr-qc/9904002} (1999).

\noindent [52] S. Chandrasekhar and V. Ferrari,  Proc. R. Soc. (London)
A {\bf 435}, 645 (1991).

\nnn [53] J. Katz and D. Lerer, Class. Quantum Grav.  {\bf 14}, 2297 (1997).

\noindent [54] A.N. Petrov, Int. J. Mod. Phys. D {\bf 4}, 451 (1995).

\noindent [55] E.T. Newman  and  T.W.J. Unti,  J. Math. Phys. {\bf 3},
891 (1962).

\end